\documentclass[twocolumn]{aastex631} 

\usepackage{enumitem} 
\setlist{nosep,after=\vspace{\baselineskip}} 
\usepackage[T1]{fontenc} 
\usepackage{placeins} 
\usepackage{graphicx,subfigure}	
\usepackage{wasysym} 

\shorttitle{Low-redshift quasar neighborhoods}
\shortauthors{Stone et al.}
\graphicspath{{./}}
\begin{document}

\title{Galaxy and Mass Assembly (GAMA): Low-redshift Quasars and Inactive Galaxies Have Similar Neighbors}

\correspondingauthor{Maria Babakhanyan Stone}
\email{mbstone12@gmail.com}

\author[0000-0002-2931-0593]{Maria B. Stone}
\affiliation{Department of Physics and Astronomy, Vesilinnantie 5, FI-20014 University of Turku, Finland}
\affiliation{Finnish Centre for Astronomy with ESO (FINCA), Vesilinnantie 5, FI-20014 University of Turku, Finland}

\author[0000-0002-7135-2842]{Clare F. Wethers}
\affiliation{Department of Space, Earth and Environment, Chalmers University of Technology, Onsala Space Observatory, 439 92, Onsala, Sweden}

\author[0000-0003-1455-7339]{Roberto de Propris}
\affiliation{Finnish Centre for Astronomy with ESO (FINCA), Vesilinnantie 5, FI-20014 University of Turku, Finland}
\affiliation{Department of Physics and Astronomy. Botswana International University of Science and Technology (BIUST), Private Bag 16, Palapye, Botswana}

\author[0000-0003-0133-7644]{Jari Kotilainen}
\affiliation{Finnish Centre for Astronomy with ESO (FINCA), Vesilinnantie 5, FI-20014 University of Turku, Finland}
\affiliation{Department of Physics and Astronomy, Vesilinnantie 5, FI-20014 University of Turku, Finland}

\author[0000-0001-9738-3594]{Nischal Acharya}
\affiliation{Donostia International Physics Center (DIPC), Paseo Manuel de Lardizabal 4, E-20018 Donostia-San Sebastian, Spain}

\author[0000-0002-4884-6756]{Benne W. Holwerda}
\affiliation{Department of Physics and Astronomy, University of Louisville, Natural Science Building 102, Louisville, KY, 40292, USA}

\author[0000-0001-5290-8940]{Jonathan Loveday}
\affiliation{Astronomy Centre, University of Sussex, Falmer, Brighton BN1 9QH, UK}

\author[0000-0001-5991-3486]{Steven Phillipps}
\affiliation{Astrophysics Group, School of Physics, University of Bristol, Tyndall Avenue, Bristol BS8 1TL, UK}

\begin{abstract}
We explore the properties of galaxies in the proximity (within a $\sim$2 Mpc radius sphere) of Type I quasars at 0.1\textless z\textless0.35, to check whether and how an active galaxy influences the properties of its neighbors.
We further compare these with the properties of neighbors around inactive galaxies of the same mass and redshift within the same volume of space, using the Galaxy and Mass Assembly (GAMA) spectroscopic survey.
Our observations reveal no significant difference in properties such as the number of neighbors, morphologies, stellar mass, star formation rates, and star formation history between the neighbors of quasars and those of the comparison sample.
This implies that quasar activity in a host galaxy does not significantly affect its neighbors (e.g. via interactions with the jets). 
Our results suggest that quasar host galaxies do not strongly differ from the average galaxy within the specified mass and redshift range.
Additionally, the implication of the relatively minor importance of the environmental effect on and from quasars is that nuclear activity is more likely triggered by internal and secular processes.
\end{abstract}


\section{Introduction} \label{sec:intro}

\defcitealias{Wethers_2022}{W22}
\defcitealias{Stone_2021}{S21}

Active galactic nuclei (AGN) are supermassive black holes (SMBHs) observed in the process of accreting matter \citep[e.g.][]{Lynden-Bell_1969}.
Although SMBHs are likely present within the centers of all massive galaxies \citep[][and references therein]{Kormendy_2013}, it is still unclear how the activity is switched on or off.
Driving gas and dust to galaxy centers to feed the SMBH is found to be difficult \citep[e.g.][]{Alexander_2012}, however numerical simulations show that gravitational torques during galaxy mergers and interactions are able to produce inflows of gas towards the galactic central regions \citep{Barnes_1991,Wurster_2013,Newton_2013,Blumenthal_2018} and feed central SMBHs, triggering AGN activity \citep{Sanders_1988,Debuhr_2011}.
Pericenter passes among galaxy pairs also correlate with peaks in both SMBH and stellar activity \citep{Gabor_2016}.\@
Secular evolution instead occurs when instabilities in galaxies drive gas inward and fuel the SMBH \citep{Shlosman_1989,Hopkins_2010}. Such secular processes may involve bar instabilities or spiral arms \citep[e.g.][]{Villforth_2019,Smethurst_2021}.
This also may trigger star formation (SF) and AGN activity.

Observational evidence regarding which one is the dominant mechanism remains ambiguous.
Some studies find no significant evidence that the hosts of AGN are more likely to be involved in mergers, compared to samples of inactive galaxies with the same luminosity and redshift \citep{Cisternas_2011,Villforth_2017,Marian_2020,Holwerda_2021}, while others make the opposite claim, particularly for IR-selected AGN \citep{Fan_2016,Goulding_2018,Zhuang_2022}.
It is possible that the merger mechanism is only important for the more luminous AGN \citep[e.g., $L_{bol}$~\textgreater 10$^{45}$~erg~s$^{-1}$,][]{Treister_2012}, or depends on other properties such as luminosity, obscuration and age \citep{Hopkins_2006_merger,Hopkins_2006_secular,Hopkins_2010}. 

As the environment may affect the properties of AGN hosts through mergers and interactions \citep[e.g.][]{Kormendy_2009,Lietzen_2011,Porqueres_2018,Wethers_2022}, so may AGN also affect the properties of their immediate neighbors \citep[e.g.][]{Gunn_1979,Silk_Rees_1998,Dashyan_2019,Martin_2020,Martin_2021,Zana_2022}, e.g. through jets up to Mpc scales \citep{Schaye_2010,Padovani_2016,Blandford_2019}.
Therefore the properties of galaxies neighboring AGN may provide information on the influence of the AGN on its neighbors, as well as on the triggering mechanism and the characteristics of AGN hosts: as an example, if most companion galaxies appear to be early-type, it would be more likely for the AGN host to be an early-type galaxy as well \citep[][]{Dressler_1980}.

It is broadly accepted that feedback from the SMBH activity affects the properties of host galaxies, such as their star formation rate \citep[SFR;][]{Sanchez_2018,Martin_Navarro_2022}.
The AGN activity, however, also influences the processes and the recycling of elements in the circumgalactic medium \citep[][]{Tumlinson_2017} and beyond, potentially influencing the properties of the galaxies around the AGN \citep{Molnar_2017,Martin_2019}.
The influence may depend on quasar host luminosity, mass, SFR, group density, redshift, satellite distance from the quasar, as well as quasar type (obscured or unobscured).\@
A recent study of an archival sample of satellite galaxies of brightest group galaxies (more likely to host or have hosted a quasar, as also shown by \citealt[hereafter \citetalias{Wethers_2022}]{Wethers_2022}) by \cite{Martin_2021} showed that galaxies were less quenched along the direction of the quasar jets, where they excavate the intragroup medium and make ram stripping less effective.

There have been claims that regions around AGN are overdense \citep{Serber_2006,Strand_2008,Zhang_2013} but these lack comparison samples around inactive galaxies. 
Most previous studies have concentrated on finding samples of companion galaxies \citep[e.g.][]{Villarroel_2012,Karhunen_2014,Yue_2019} to measure the merger rate around AGN.
\citet{Coldwell_2003,Coldwell_2006} claim that there is an excess of star-forming disk galaxies within 1 Mpc of AGN.
Modest evidence was further found for star-forming neighbors of low-redshift quasars \citep[hereafter \citetalias{Stone_2021}]{Bettoni_2017,Bettoni_2022,Stone_2021}.

In order to properly compare the environments of quasar host galaxies with those of currently inactive objects, we need large spectroscopic samples, allowing us to select both quasars and normal galaxies in the same manner and therefore isolate the influence of the environment on quasars and vice versa.
Galaxy and Mass Assembly \citep[GAMA][]{Driver_2011,Liske_2015,Baldry_2018,Driver_2022} is a highly complete spectroscopic survey for redshifts of z\textless 0.4.
Additionally, the GAMA dataset includes multi-wavelength photometry (over 22 bands) and rich ancillary information (derived quantities such as stellar mass, specific SF).

The GAMA survey is ideally suited for statistical studies of neighborhoods of low-redshift galaxies at the sub-Mpc scale, since it covers a sufficiently large area and its redshift measurements are reliable, spatially uniform, highly complete and extend to the galaxies with faint fluxes.
The observational methodology and instrument configurations of GAMA survey provide high target density, thus making it highly complete even for close pairs \citep{DePropris_2014,Robotham_2014}.
In particular, GAMA survey is designed to have multiple passes to the same region, thus it is not affected by fibre placement restrictions, which consequently ensures the redshift completeness of close pairs \citep[][]{Driver_2011,Robotham_2014,Liske_2015}.
The "close pair" definition is based on a magnitude-redshift space-based samples to ensure completeness, as shown in GAMA survey-based study of close pair fractions at low redshift \citep[][]{DePropris_2014}.
The survey extends down to a magnitude limit of Sloan Digital Sky Survey (SDSS) $r$ \textless 19.8 mag.
It is $r$-selected, so even sources with dust obscuration are less likely to be missed by the survey.

\citetalias{Wethers_2022} exploited the GAMA data, showing that quasar host galaxies, compared to a sample of inactive galaxies matched in mass and redshift, did not occupy significantly different environments.\@ 
The high completeness and extensive information derived for GAMA targets allows us to consider the opposite question - whether the properties of galaxies in the neighborhood of quasars are affected by the proximity of quasar host galaxies.

Here we use GAMA data to expand on the work by \citet{Stone_2021} and \cite{Bettoni_2017,Bettoni_2022}, on low-redshift (z\textless 0.5) quasar environments in the Sloan Digital Sky Survey \citep[SDSS][]{Ahumada_2020} Stripe82 region \citep{Annis_2014}.
In these studies, long slits were placed on galaxies within several hundred kpc of projected distance of quasars, to identify physically associated companion galaxies (at the same redshift) and to determine their properties.
Similarly, here, we identify spectroscopically-confirmed companion galaxies contained within a fixed comoving volume (radius $\sim$ 2 Mpc) for each quasar from the independent sample of quasars defined in \citetalias{Wethers_2022}, and compare the properties of these neighboring galaxies against a comparison sample of neighboring galaxies around inactive galaxies whose properties (stellar mass and redshift) are matched to those of the quasar hosts.

The outline of the paper is as follows.\@
The quasar sample, the comparison galaxy samples, and the identification of neighboring galaxies from the GAMA survey are described in \S~\ref{sec:data}.\@
The results of the comparison between the neighbors around quasars and inactive galaxies are presented in \S~\ref{sec:results} and discussed in \S~\ref{sec:discussion}.\@
We summarize in \S~\ref{sec:conclusions}.\@
The analysis is performed assuming the flat $\Lambda$ cold dark matter ($\Lambda$CDM) cosmology model with the following parameters: the matter density parameter $\Omega_M$=0.3, the cosmological constant $\Omega_{\Lambda}$=0.7, the Hubble constant H$_0$=70 km s$^{-1}$ Mpc$^{-1}$ \citep[][]{Planck_VI}.

\section{DATA} \label{sec:data}

This work is based on the whole sample of 205 Type I quasars at 0.1\textless z\textless 0.35 as defined in \citetalias{Wethers_2022}.
In short, these quasars were selected from the version 4 of the Large Quasar Astrometric Catalog \citep[LQAC4;][]{Gattano_2018}, constrained to the GAMA equatorial regions, and restricted to be above the GAMA survey depth (\textit{r}\textless 19.8).

\begin{figure*}[hbt]
    \gridline{\fig{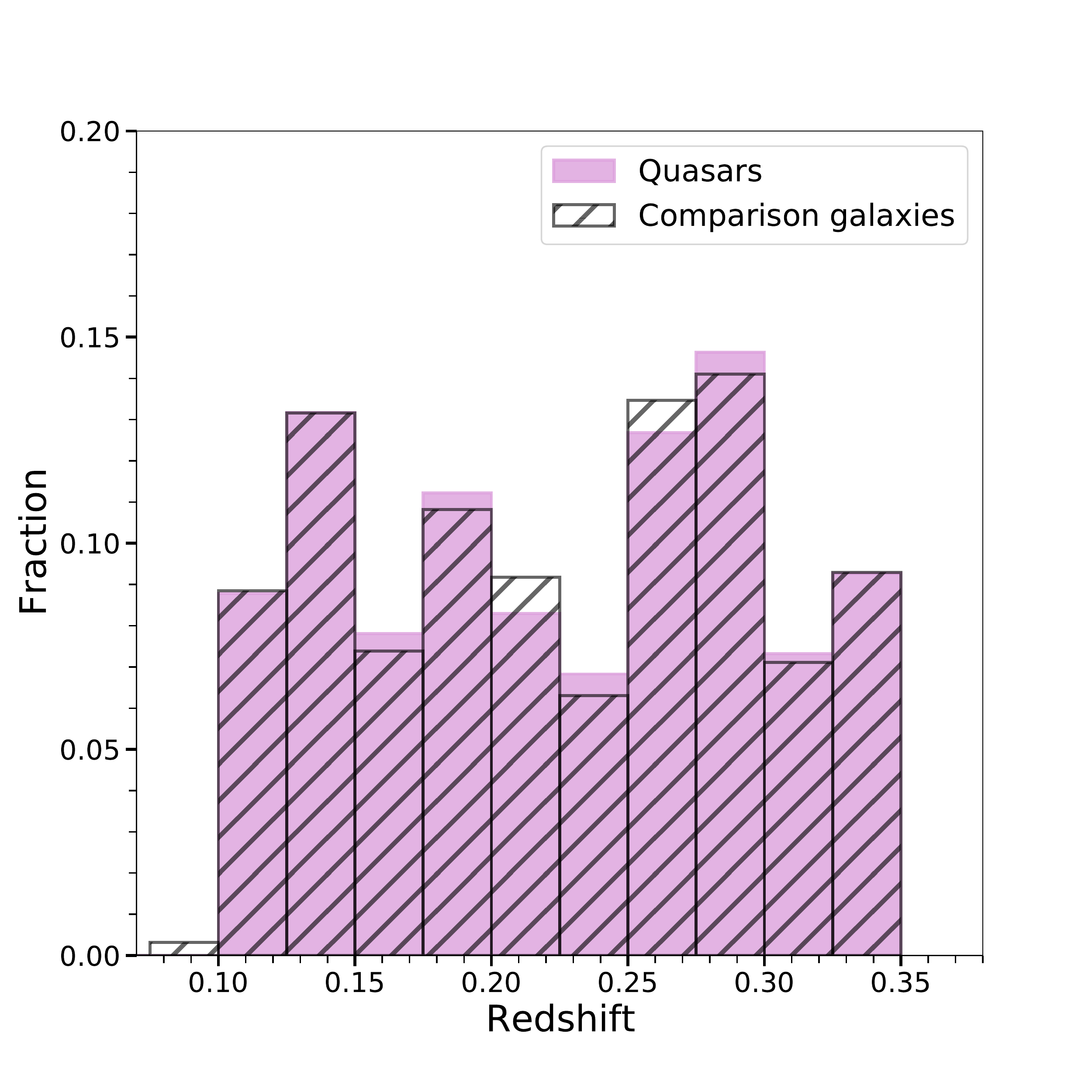}{0.47\textwidth}{(a)}
            \fig{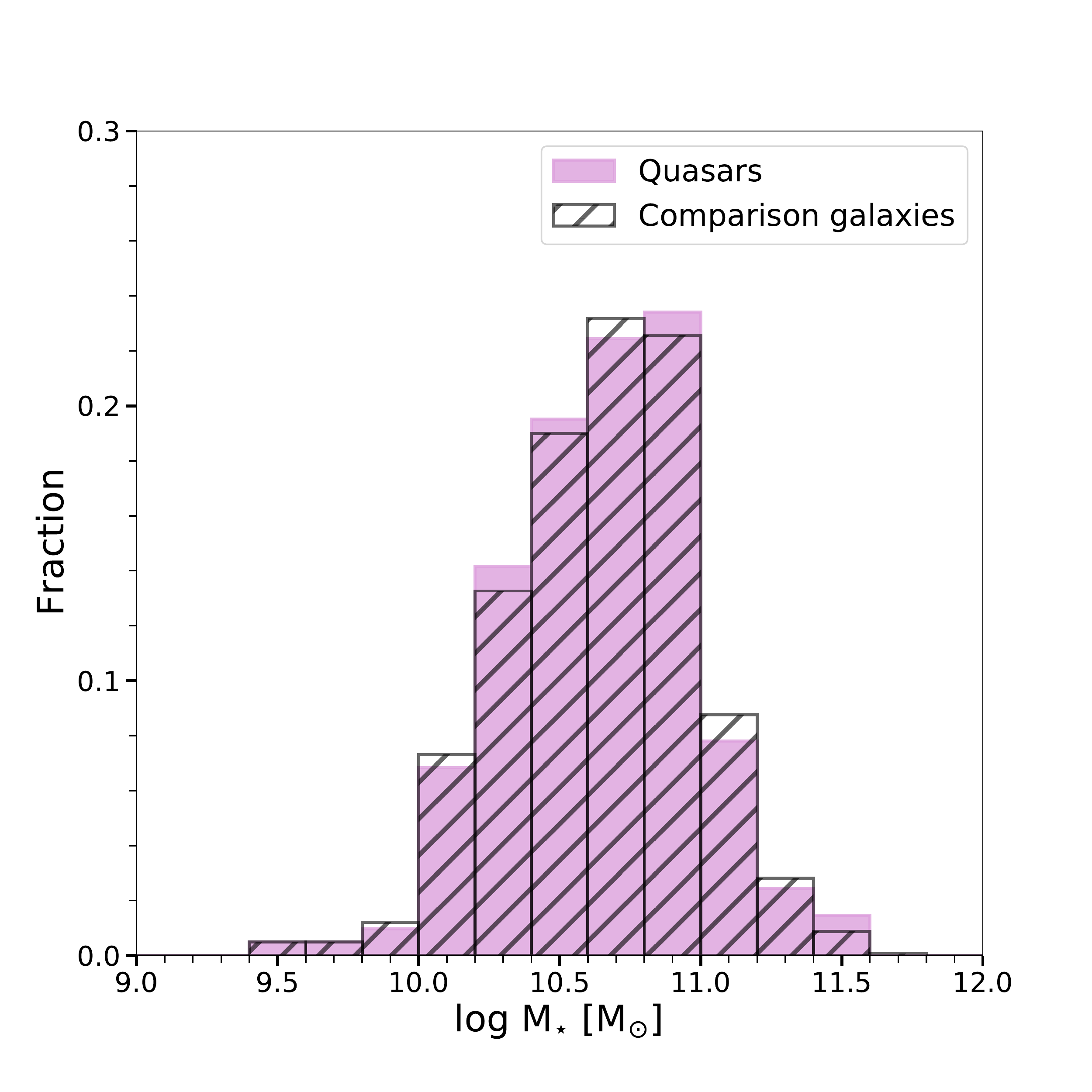}{0.47\textwidth}{(b)}
          }
    \caption{Distributions of (a) redshift and (b) stellar mass for quasars and for the matched inactive galaxies.}
    \label{fig:comparison_samples}
\end{figure*}

We next used GAMA to select a comparison sample of inactive galaxies matched in redshift and stellar mass to our sample of quasar hosts, following the method used in \citetalias{Wethers_2022}.
We created 200 such `Monte-Carlo' realizations, which is sufficiently high to have robust statistical comparison.
Each realization contains 205 matched inactive galaxies, matching the number of quasars in our sample.
The redshift and stellar mass distributions of the quasar sample and the matched inactive galaxies (from all realizations combined) show that the selected comparison galaxies are closely matched to quasars (Fig.~\ref{fig:comparison_samples}).

\begin{figure*}[hbt]
    \gridline{\fig{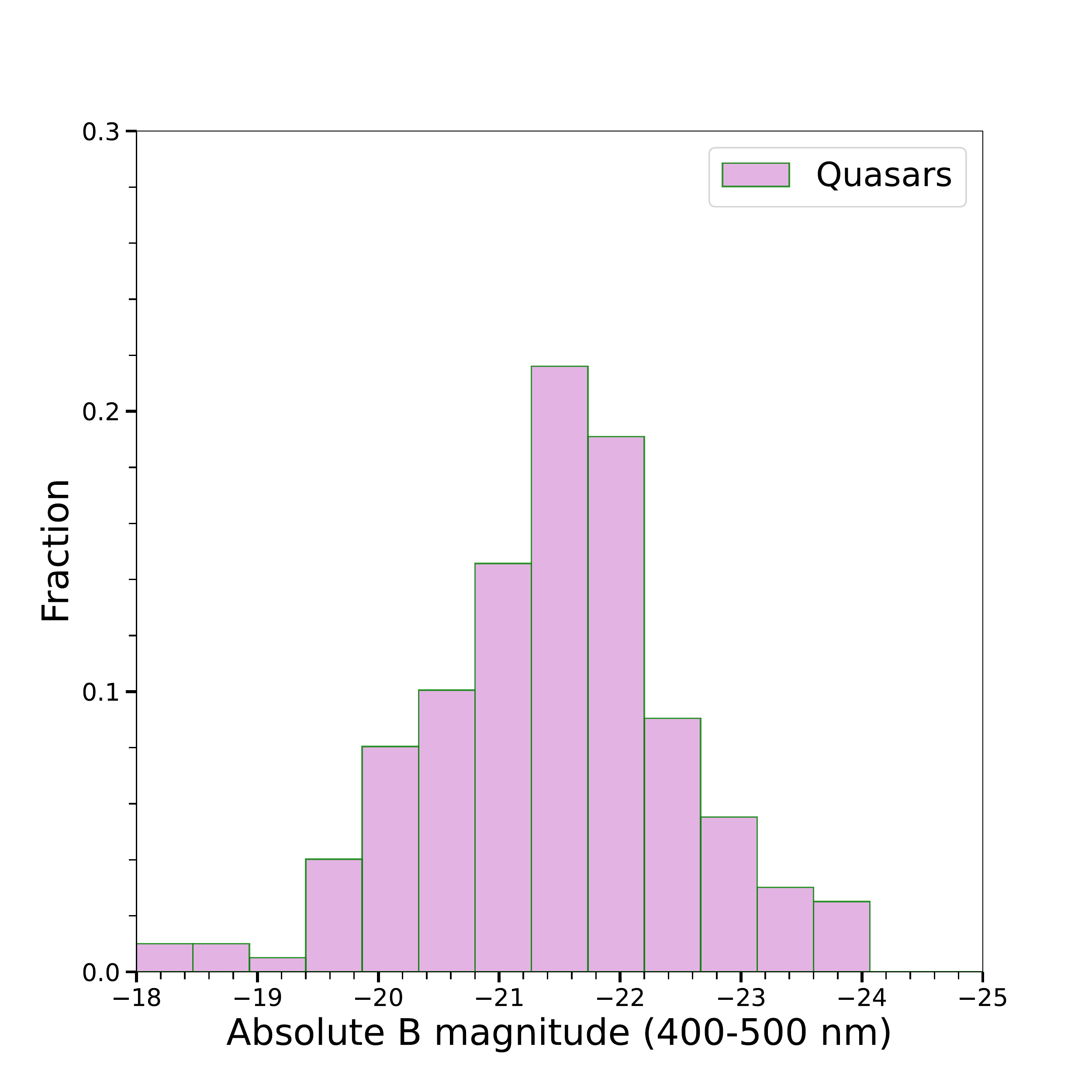}{0.47\textwidth}{(a)}
              \fig{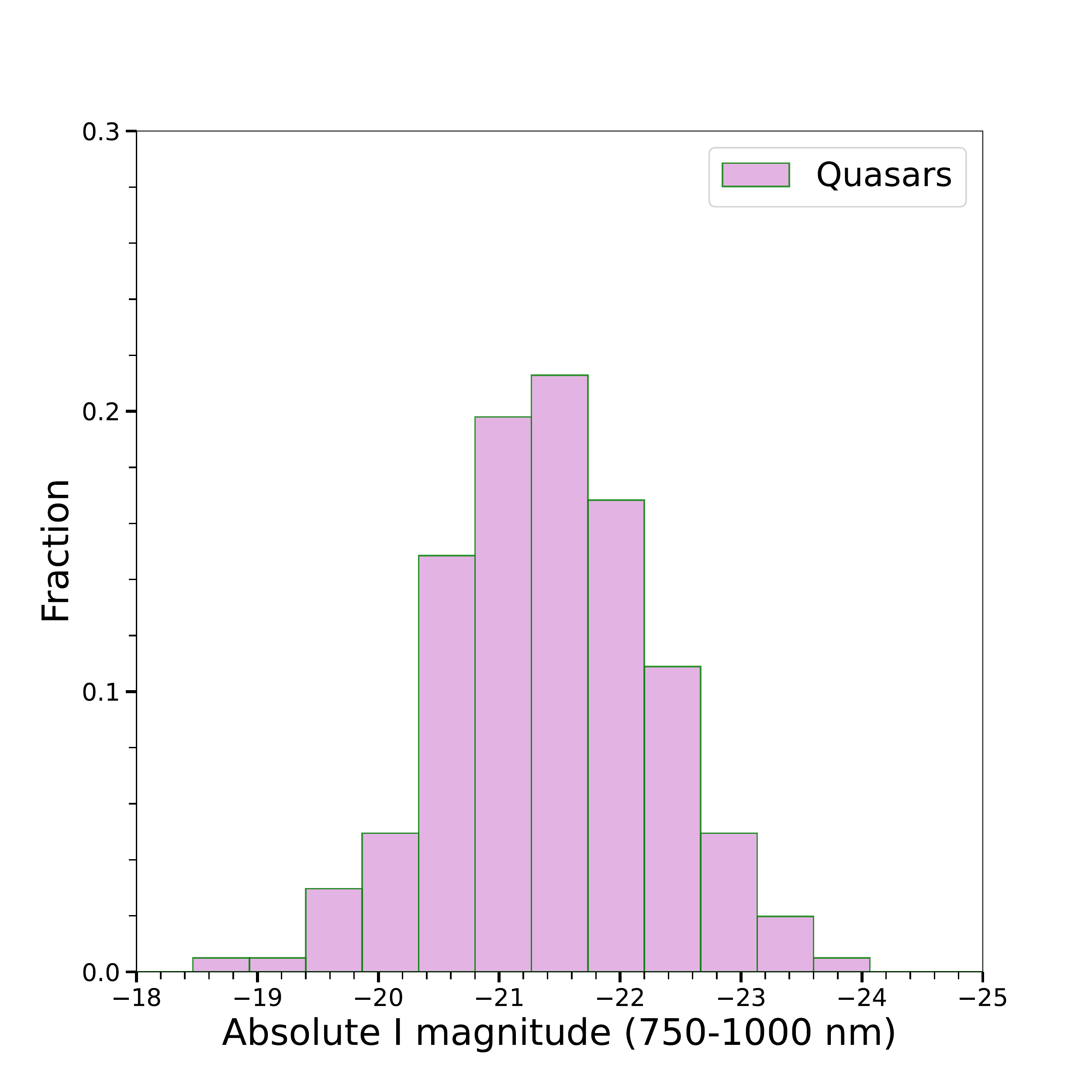}{0.47\textwidth}{(b)}
              }
    \caption{Distributions of absolute magnitudes in $B$ and $I$ for LQAC4 quasars in this sample.}
    \label{fig:luminosities}
\end{figure*}

The luminosity distribution of the quasar sample is shown in Fig.~\ref{fig:luminosities}, using the available absolute magnitudes reported in the LQAC4 catalog \citep[][]{Gattano_2018,Ochsenbein_2000}.
Note that the luminosities here include contributions from the nuclear activity as well as from the host galaxy.
The median value for absolute B magnitude is -21.50$^{-20.81}_{-22.08}$ (the uncertainty is reported as lower and upper quartiles).
The median value for absolute I magnitude is -21.42$^{-20.84}_{-21.98}$).
As expected, our sample based on LQAC4 and GAMA consists of relatively low luminosity quasars, i.e., the typical objects occupying a specified volume of space.

\begin{figure}[hbtp]
    \centering
    \includegraphics[width=0.47\textwidth]{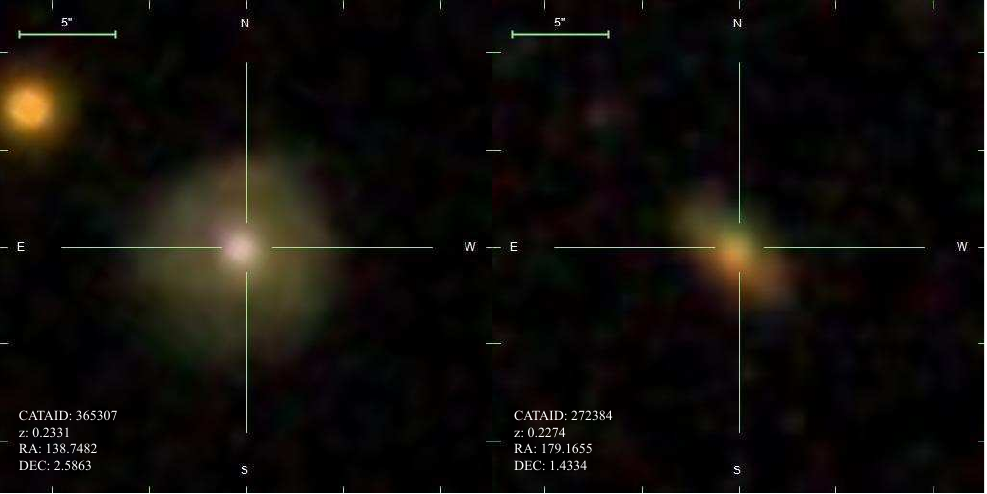}
    \includegraphics[width=0.47\textwidth]{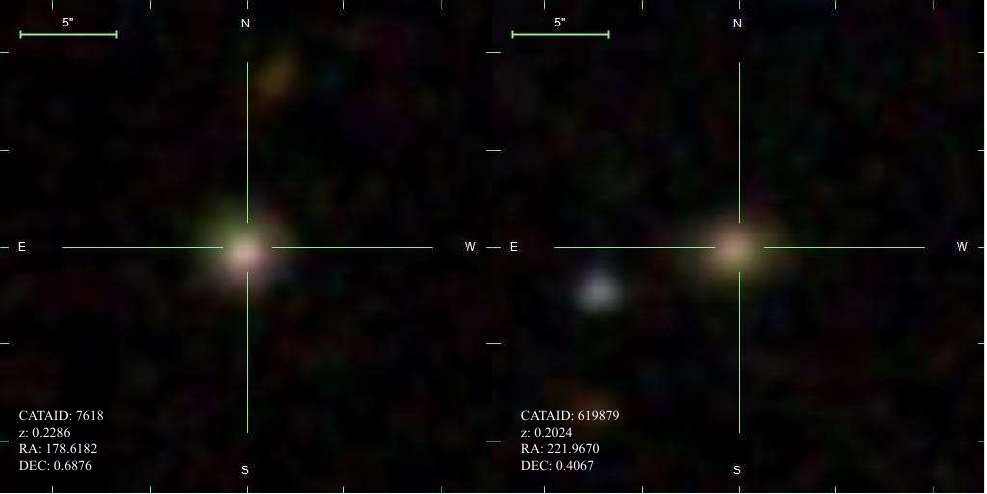}
    \includegraphics[width=0.47\textwidth]{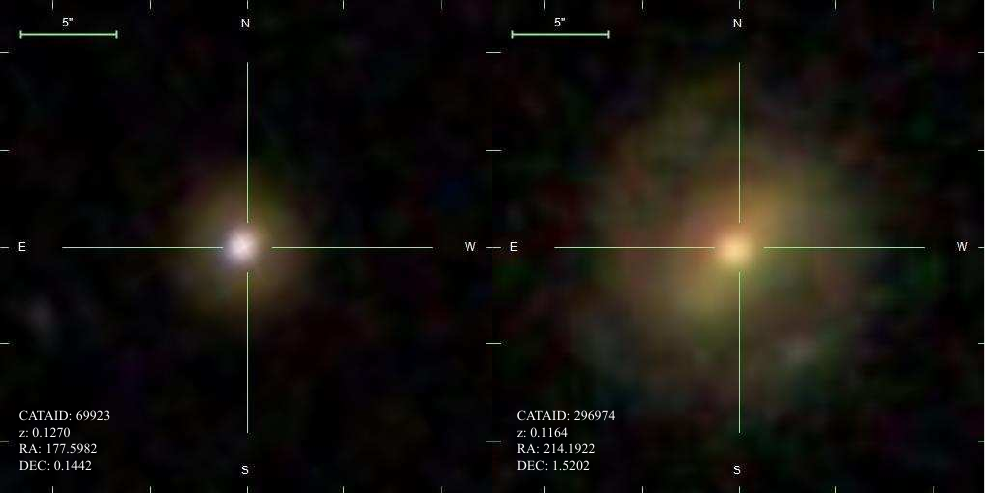}
    \includegraphics[width=0.47\textwidth]{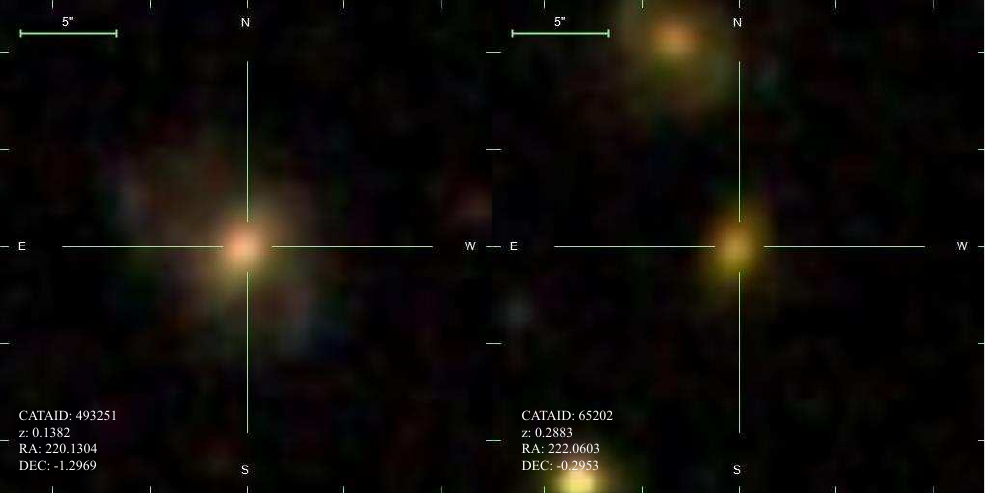}
    \caption{Quasar host galaxy examples in our sample.}
    \label{fig:quasar_hosts_images}
\end{figure}

\begin{figure}[hbtp]
    \centering
    \includegraphics[width=0.47\textwidth]{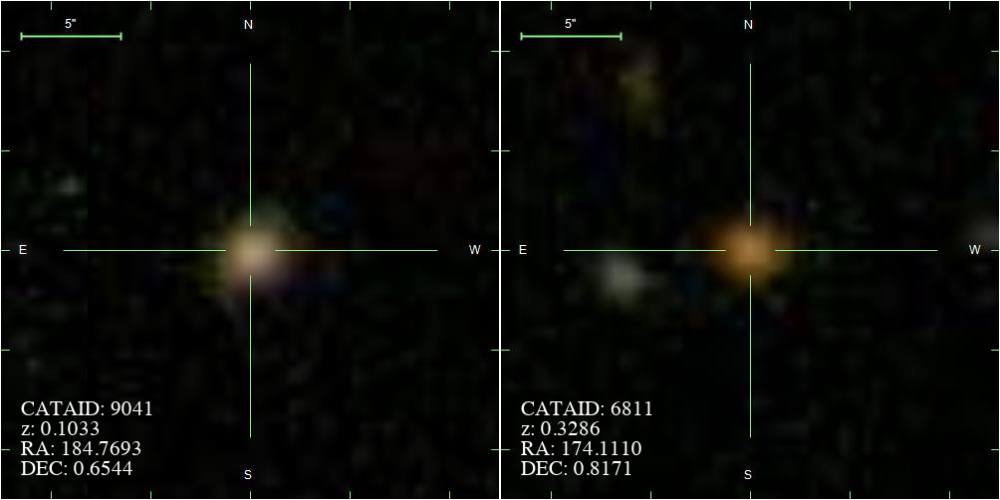}
    \includegraphics[width=0.47\textwidth]{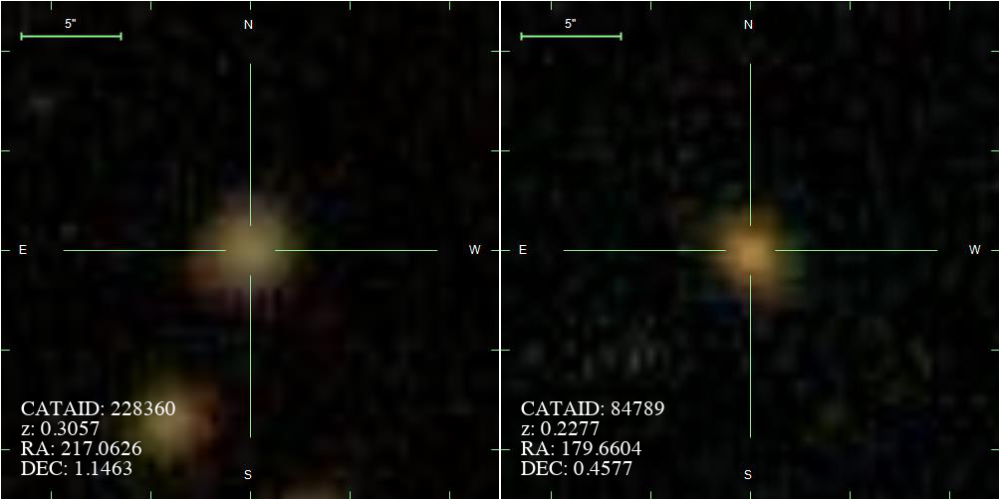}
    \includegraphics[width=0.47\textwidth]{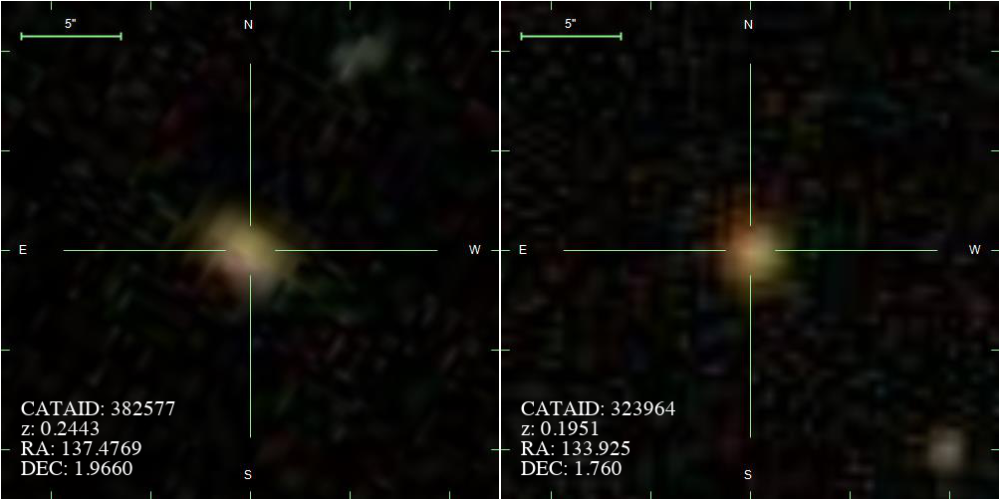}
    \includegraphics[width=0.47\textwidth]{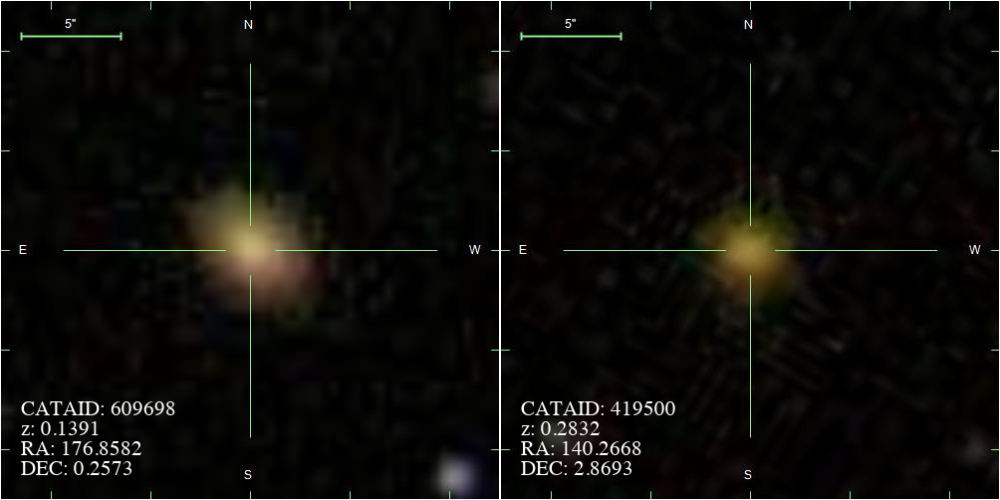}
    \caption{Comparison galaxy examples.}
    \label{fig:compgal_images}
\end{figure}

The median log(SFR) estimate for quasars (comparison galaxies) is 0.90 (-0.12) M$_{\odot}$ yr$^{-1}$, however the contribution of AGN is not accounted for by the version of MAGPHYS used in the GAMA survey and these estimates have to be
taken under advisement.
The range of log(SFR) for quasars is -0.36$-$2.22 M$_{\odot}$ yr$^{-1}$, while for comparison galaxies, -3.64$-$2.22 M$_{\odot}$ yr$^{-1}$.
For the purpose of comparison, we show few example images of quasar host galaxies and comparison galaxies in Fig.~\ref{fig:quasar_hosts_images} and \ref{fig:compgal_images} \citep{Ahn_2014}.

Next, to find the neighbors for each quasar, we select galaxies within a fixed comoving sphere of radius (R$\sim$1.8 Mpc) using the positions and redshifts provided by the latest GAMA catalog \citep[{\fontfamily{qcr}\selectfont
SpecCat} data management unit (DMU),][]{Baldry_2014,Baldry_2018,Liske_2015}.
This radius corresponds to the characteristic radius R$_{200}$ \citep[][]{Carlberg_1996,Carlberg_1997} for a system with an internal velocity dispersion of 700 km s$^{-1}$, which is equivalent to a relatively poor cluster or a rich group, i.e. the typical environment in which most galaxies tend to reside, including the AGN host galaxies in \citetalias{Wethers_2022}.
A similar comoving volume was surveyed around each of the comparison galaxies.
The GAMA survey redshifts have median velocity uncertainty of 33 km s$^{-1}$ \citep{Baldry_2014}.

We use this sample to investigate the close environment of AGN hosts and compare it to that of inactive galaxies.
Additionally, we use derived quantities from the GAMA survey to compare the properties of neighbors for both AGN host galaxies and matched inactive galaxies by performing the two-sample Kolmogorov-Smirnov (KS) test, as in \citetalias{Wethers_2022}, implemented with the {\fontfamily{qcs}\selectfont
scipy.stats.ks\_2samp} routine.
Table~\ref{table:properties} shows the derived quantities used and the appropriate GAMA DMU employed, with reference to the published description in each case.
Since the KS test is not as sensitive to the wings of the distribution, a two-sample Anderson-Darling (AD) test was also performed using the {\fontfamily{qcs}\selectfont
scipy.stats.anderson\_ksamp} package.

\begin{table*}[hbt!]
	\centering
	\caption{}
	\vspace{-3mm}
	\begin{center}
	    KS and AD Test Results
	\end{center}
	\hspace{-1.5cm}
	\begin{tabular}{ lllcll }
	    \hline
	    \hline
	    Property & Units & Statistic & $p$-value & Figure & Ref.\\
	     &  & D$_{KS}$ (A$_{k}$)  & $p_{KS}$ ($p_{AD}$) &  & \\
	    \hline
	\textbf{DISTANCE} & & & & & \\   
	Comoving separation from quasar (comparison galaxy)         & Mpc & 0.06 (-0.5) & 0.84 (0.25) & Fig.~\ref{fig:comoD_sep_hist} & \\
	\textbf{MORPHOLOGY} from {\fontfamily{qcr}\selectfont S\'ersic photometry} DMU & & & & & 1,4,5 \\   
	S\'ersic index (n)         &...& 0.15 (-0.4) & 0.12 (0.25) & Fig.~\ref{fig:sersic_index_results} & \\
	\textbf{COLORS} from {\fontfamily{qcr}\selectfont LambdarPhotometry} DMU & & & & & 2,6\\   
	SDSS $u-r$ &...& 0.08 (-0.6) & 0.80 (0.25) & Fig.~\ref{fig:color_u_r}a & \\
	SDSS $g-r$ &...& 0.10 (-0.5) & 0.54 (0.25) & Fig.~\ref{fig:color_g_r}a & \\
	\textit{GALEX} $NUV-$ SDSS $r$  &...& 0.09 (-0.7) & 0.77 (0.25) & Fig.~\ref{fig:color_nuv_r}a & \\
	\textbf{SFH \& PHYSICAL PARAMETERS} from {\fontfamily{qcr}\selectfont MagPhys} DMU & & & & & 1,3 \\
	SFR          & M$_{\odot}$ yr$^{-1}$ & 0.06 (-0.9) & 0.96 (0.25) & Fig.~\ref{fig:SFR_results}a & \\
	SSFR         & yr$^{-1}$             & 0.07 (-0.7) & 0.89 (0.25) & Fig.~\ref{fig:SFR_results}b & \\
	SFR$_{ave}$ over the last $10^{7}$ yr        & M$_{\odot}$ yr$^{-1}$ & 0.07 (-0.2) & 0.88 (0.25) & Fig.~\ref{fig:sfr17_results}a & \\
	SFR$_{ave}$ over the last $10^{8}$ yr        & M$_{\odot}$ yr$^{-1}$ & 0.07 (-0.7) & 0.86 (0.25) & Fig.~\ref{fig:sfr17_results}b & \\
	SFR$_{ave}$ over the last $10^{9}$ yr        & M$_{\odot}$ yr$^{-1}$ & 0.09 (-0.5) & 0.67 (0.25) & Fig.~\ref{fig:sfr17_results}c & \\
	SFR$_{ave}$ over the last $2\times10^{9}$ yr        & M$_{\odot}$ yr$^{-1}$ & 0.11 (-0.1) & 0.37 (0.25) & Fig.~\ref{fig:sfr17_results}d & \\
	SF timescale   &   Gyr$^{-1}$ & 0.08 (-0.8) & 0.73 (0.25) & Fig.~\ref{fig:SF_timescale_results}a & \\
	Time since the last burst of SF ended & dex(yr)           & 0.09 (-0.6) & 0.64 (0.25) & Fig.~\ref{fig:SF_timescale_results}b & \\
	Total M$_{\star}$ ever formed (integral of the SFR) & M$_{\odot}$           & 0.12 (0.1) & 0.32 (0.25) & Fig.~\ref{fig:mass_results}a & \\
	Total mass of dust         & M$_{\odot}$             & 0.05 (-0.9) & 0.99 (0.25) & Fig.~\ref{fig:mass_results}b & \\
	Fraction of mass formed in bursts over the last $10^{7}$ yr &...& 0.04 (18.9) & 0.99 (0.001) & Fig.~\ref{fig:fb17_results}a & \\
	Fraction of mass formed in bursts over the last $10^{8}$ yr &...& 0.04 (6.0) & 0.99 (0.001) & Fig.~\ref{fig:fb17_results}b & \\
	Fraction of mass formed in bursts over the last $10^{9}$ yr &...& 0.08 (-0.3) & 0.84 (0.25) & Fig.~\ref{fig:fb17_results}c & \\
	Fraction of mass formed in bursts over the last $2\times10^{9}$ yr &...& 0.15 (0.4) & 0.12 (0.22) & Fig.~\ref{fig:fb17_results}d & \\
	Age of the oldest stars in the galaxy & dex(yr)               & 0.07 (-0.9) & 0.88 (0.25) & Fig.~\ref{fig:age_results}a & \\
	Metallicity  & Z$_{\odot}$           & 0.12 (-0.2) & 0.33 (0.25) & Fig.~\ref{fig:age_results}b & \\
	    \hline 
	\vspace{0.5mm}
	\end{tabular}
	{\raggedright \textbf{Notes.} KS and AD test comparisons between the spectroscopically confirmed neighbors of quasars and matched inactive galaxies.
	D$_{KS}$ is the calculated KS test statistic.
	A$_{k}$ is the calculated AD test statistic for k$=$2 samples.
	The AD test statistic and $p$-values are in parentheses.
	The $p$-values of the AD test are capped at 25\%.
	For {\fontfamily{qcr}\selectfont MagPhys} DMU we used the latest internal version.
	\par
	}
	{\raggedright \textbf{References.}
	(1) \citet{Baldry_2018};
	(2) \citet{Driver_2016};
	(3) \citet{Driver_2018};
	(4) \citet{Hill_2011};
	(5) \citet{Kelvin_2012};
	(6) \citet{Wright_2016}.
	\par
	}
	\label{table:properties}
\end{table*}

\subsection{GAMA ancillary products}

The various galaxy properties have been derived by the GAMA collaboration as ancillary information.
These data are included in the survey as {\fontfamily{qcr}\selectfont S\'ersic photometry}, {\fontfamily{qcr}\selectfont LambdarPhotometry}, and {\fontfamily{qcr}\selectfont MagPhys} DMUs.
In this section we briefly describe how GAMA derived these properties.
We refer the reader to the references for further details.

The colors of neighbors are computed using the AB system magnitudes supplied by the GAMA {\fontfamily{qcr}\selectfont LambdarPhotometry} DMU \citep{Wright_2016,Driver_2016}.
The deblended matched aperture photometry is calculated using the \textsc{lambdar} code ("lambda adaptive multi-band deblending algorithm in R").
We used SDSS $u-r$, $g-r$, and \textit{GALEX} NUV$-$SDSS $r$ colors \citep[\textit{GALaxy Evolution eXplorer} satellite]{Martin_2005}.
We exploited the following individual filter catalogs: LambdarSDSSuv01, LambdarSDSSgv01, LambdarSDSSrv01, and LambdarGALEXNUVv01 to retrieve the reported AB magnitudes of final deblended flux and their associated errors.

To describe the morphologies, the S\'ersic indices reported are surface brightness distribution fits to a single component S\'ersic profile \citep{Sersic_1968} in the SDSS $r$ band \citep[{\fontfamily{qcr}\selectfont S\'ersic photometry} DMU, version 9;][]{Hill_2011,Kelvin_2012,Baldry_2018}.
The S\'ersic index value and its associated error are retrieved from the SersicCatSDSS catalog.
The fit is produced by the SIGMA tool \citep["structural investigation of galaxies via model analysis", v1,][]{Kelvin_2012} based on the GALFIT two-dimensional fitting algorithm \citep{Peng_2002,Peng_2010}.
The code performs a non-linear least-squares fit and calculates the goodness of fit using the normalized $\chi^{2}$ computation technique, estimating the uncertainty in each pixel based on its Poisson error.

This study makes use of the physical stellar population parameters for galaxies provided by the MagPhys table within the {\fontfamily{qcr}\selectfont MagPhys} spectral energy distribution (SED) DMU \citep[][latest internal version]{Baldry_2018,Driver_2018}.
The parameters are derived through the execution of the \textsc{magphys} code (multi-wavelength analysis of galaxy physical properties) for fitting the SED of a galaxy \citep{daCunha_2008,daCunha_2011}.
\textsc{magphys} assumes energy conservation, relies on the \citet{Bruzual_2003} models, does not include AGN emission, and is based on Bayesian method.
\textsc{magphys} SED fitting results are in general agreement with similar codes in the literature \citep{Hayward_2015,Hunt_2019}.
It was applied to the 21-band GAMA photometry catalog \citep[LAMBDARCatv01][]{Wright_2016,Driver_2018} spanning far-ultraviolet (FUV) to far-infrared (FIR) wavelength regime.

\begin{figure*}[ht]
    \centering
    \includegraphics[width=0.82\textwidth]{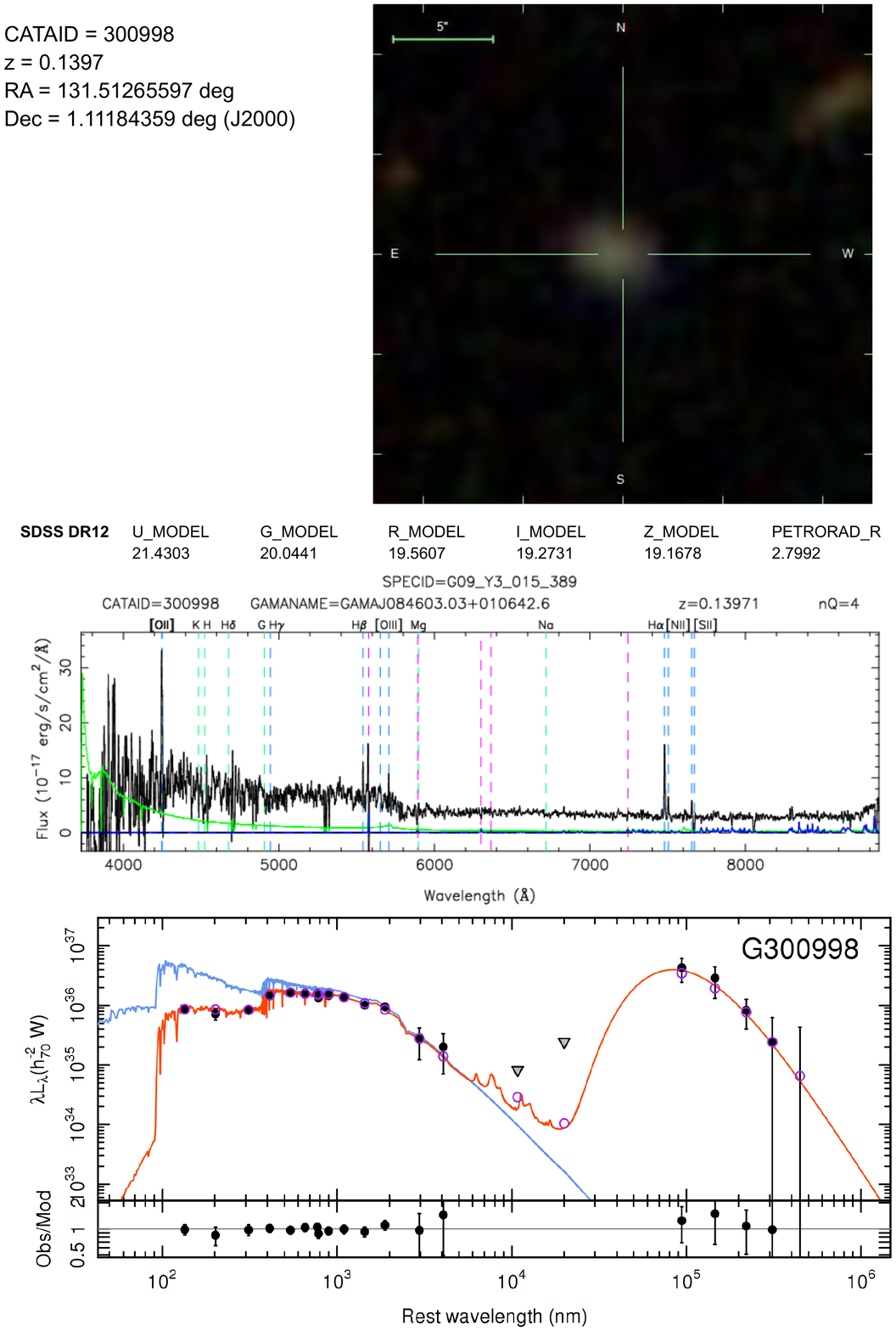}
    \caption{ Quasar neighboring galaxy example.
    Top: Postage stamp image of GAMA galaxy 300998 (GAMA Survey Image Viewer tool). 
    Middle: Spectrum (black line) and error (green line). 
    Bottom: MAGPHYS fit (orange line) to data (filled dots and gray triangles for upper limits), with residuals underneath. 
    Open magenta circles show the predicted photometry from the model. The blue line is a MAGPHYS model with no internal dust correction.
    }
    \label{fig:examplecompanion}
\end{figure*}

We focus on the galaxy-wide physical parameters pertaining to the stars: stellar mass estimates (separate from the stellar mass estimates from \citet{Taylor_2011}, see \citet{Baldry_2018} for a discussion comparing the two), SFR estimates over different timescales, specific SFR (sSFR), metallicity, age of the oldest stars in the galaxy, SF timescale, fraction of stellar mass formed in bursts over various timescales.
Fig.~\ref{fig:examplecompanion} shows the results of this procedure for one example companion galaxy and its SDSS image \citep{Ahn_2014}.

\textsc{magphys} estimates the parameters of stellar populations (age, metallicity, SF metrics) and the interstellar medium by comparing multi-band photometry against a large library of stellar templates and dust emission templates in the mid and far-infrared.
This returns best fit and median estimates for the sought parameters.
In this work we used the best fit values for the SFR, sSFR, stellar mass and dust mass, while for the rest of the SF history (SFH) and physical parameters we used the median values.

The fitting result depends on the observational data provided to the code as an input, which includes the observed fluxes and the associated observational uncertainties.
Since the GAMA fluxes range from UV to FIR, the SED fitting estimates are reliable and well-constrained.
For each parameter, the median of the resultant probability density function is considered the best estimate of that parameter.
Furthermore, the confidence interval is the 16th-84th percentile range \citep{daCunha_2008, Baldry_2018}.
For properties produced by the \textsc{magphys} SED fitting code, we also report the uncertainties using the 16-50-84 percentile method. 
We compute for each neighbor the difference between the 50th (aka median) and 16th percentiles, and then take the median of these differences for all neighbors to be the typical lower uncertainty bound ($\widetilde{\Delta}_{16}$).
For the typical upper uncertainty bound, the same calculation is done using the difference between the 84th and 50th percentiles ($\widetilde{\Delta}_{84}$).

\vspace{\baselineskip}
\section{RESULTS} \label{sec:results}

\begin{figure}
    \centering
    \includegraphics[width=0.44\textwidth]{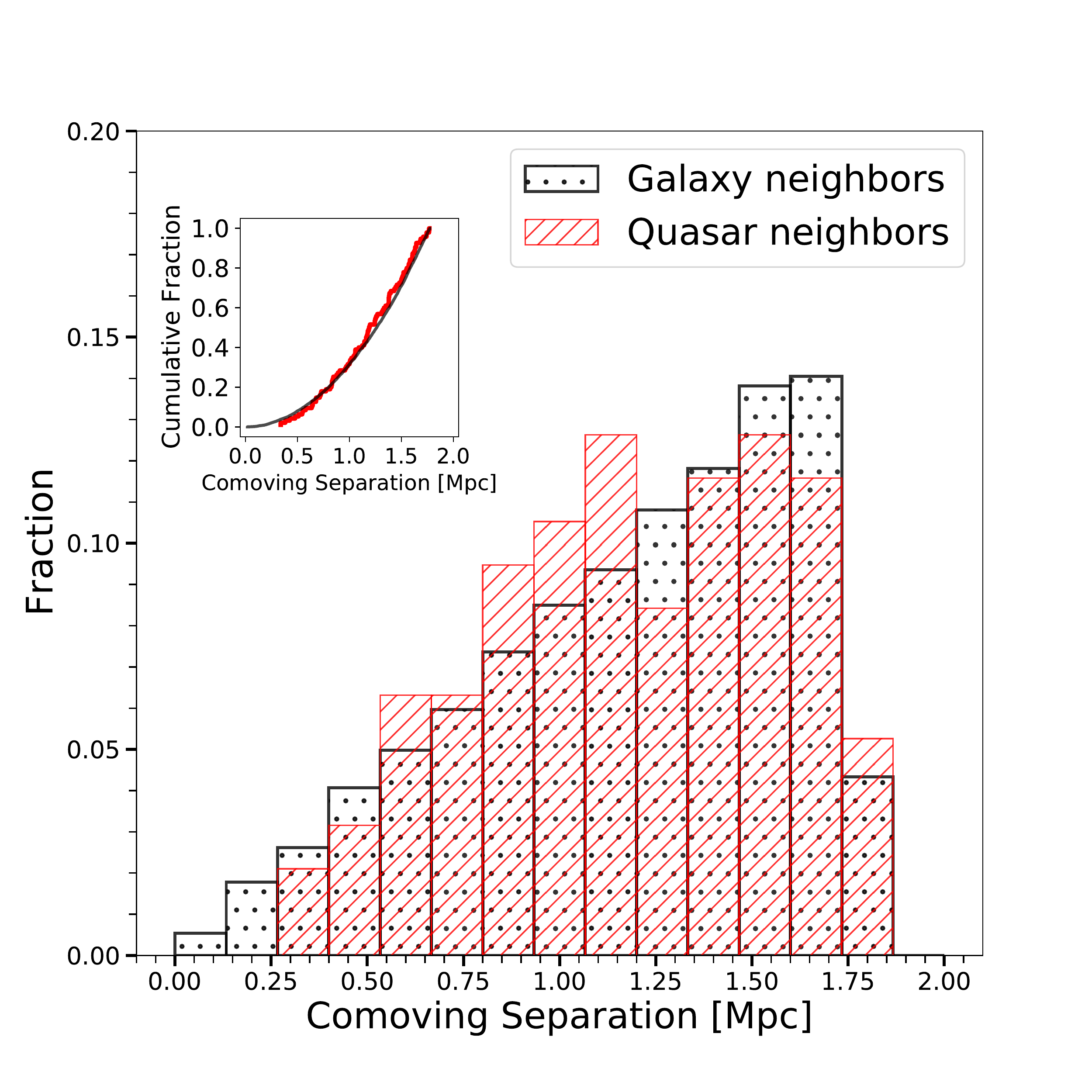}
    \caption{Comoving separation distributions of neighboring galaxies (quasar neighbors in red hatched bars, comparison galaxy neighbors in black dotted bars).
    The inset shows the empirical cumulative distribution function \citep[eCDF][]{Seabold_2010}, quasar neighbors in red solid line, and comparison galaxy neighbors in black.}
    \label{fig:comoD_sep_hist}
\end{figure}

Since mergers are more likely in denser environments and SF bursts accompany simulated interactions \citep[][]{DiMatteo_2005,Hopkins_2006_merger,Blandford_2019}, this study looks at the number counts and physical properties of galaxies in the vicinity of quasars (and matched inactive galaxies), checking whether the observational data are consistent with the merger scenario.
We first consider the counts of neighbors contained within a sphere of radius R$\sim$1.8 Mpc for AGN host galaxies and the average counts of neighbors for galaxies in the comparison sample of matched inactive galaxies.
For quasar host galaxies, 33.6\% (69 out of 205) had at least one neighbor compared to 31.7$\pm$3.3\% of the matched comparison galaxies on average. 
When considering only objects with neighbors, we find that for quasar host galaxies 68.1$\pm$9.9\% had one neighbor, 26.0$\pm$6.1\% had two, and 5.7$\pm$2.8\% had three; while for the matched inactive galaxies respectively 77.6$\pm$5.3\%, 17.8$\pm$4.9\%, and 3.5$\pm$2.1\%. 
We find no significant difference in the fractions of neighbors for AGN hosts and inactive galaxies as would be expected if AGN hosts occupied denser environments.
Additionally, Fig.~\ref{fig:comoD_sep_hist} shows the distributions of comoving separations for the neighboring galaxies of quasars and matched inactive galaxies.
The comoving separations were derived following \citet{Lindsay_2014} method.
The KS test results in no significant difference between the two populations (Table~\ref{table:properties}), unlike what would be expected if AGNs were preferentially triggered by mergers.
Therefore this suggests that there is no strong merger-AGN connection.
For reasons of clarity and brevity all subsequent figures in this section are shown in the Appendix; Table~\ref{table:properties} points to these figures.

We then compare the distributions of several derived parameters for neighbors of AGN and our comparison sample of inactive galaxies. 
Fig.~\ref{fig:sersic_index_results} shows the distribution of S\'ersic indices for neighbors of AGN hosts and for those of inactive galaxies. 
The mean values of both populations are similar (n$=$2.4$\pm$1.7 for quasar neighbors and n$=$2.6$\pm$2.6 for comparison galaxy neighbors).
The KS test does not reveal any significant statistical difference between the two samples.

Colors are shown here for comparison with other studies, although more detailed information on the properties of stellar populations is provided by GAMA's estimates of SFH. 
We have considered typical colors sensitive to age and metallicity such as $NUV-r$, $u-r$ and $g-r$ (in AB system of magnitudes). 
Again there are no significant statistical differences between the distribution of galaxy colors for neighbors of AGN hosts and those of inactive galaxies.
The color-magnitude plots show that the quasar neighbors and comparison galaxy neighbors populate the same areas.

Information on SFRs; sSFRs; stellar masses; SFRs averaged over 10, 100, 1000, and 2000 Myr; as well as the cumulative age of stars; amount of stellar mass formed over time; and metallicities are derived by GAMA using \textsc{magphys} code \citep{daCunha_2008,daCunha_2011}. 
KS tests in no case indicate any significant difference in the properties of galaxies that are AGN neighbors or neighbors of inactive galaxies (Table~\ref{table:properties}).

Since the KS tests are sensitive to the bulk of the distributions and less to the wings, additionally an alternative test, the two-sample AD test \citep{Scholz_1987} of similarity was used.
The AD test agreed with the KS test in all but two properties. 
The AD test for the fraction of mass formed in bursts over the last 10 Myr and 100 Myr showed marginal difference.

\subsection{Star Formation and Environment}
\begin{figure*}
   \centering
   \subfigure[]{\includegraphics[width=0.45\textwidth]{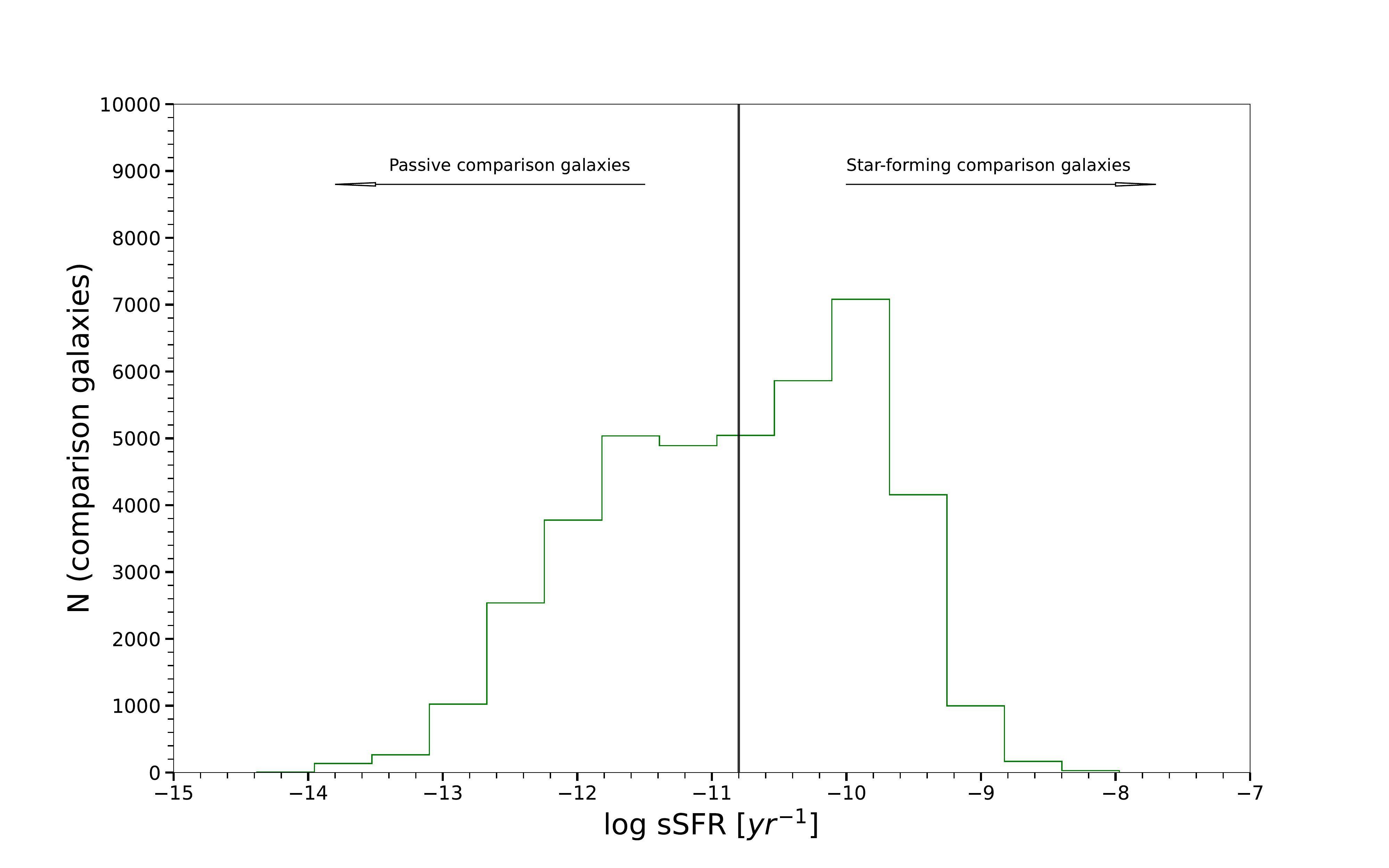}}
   \subfigure[]{\includegraphics[width=0.45\textwidth]{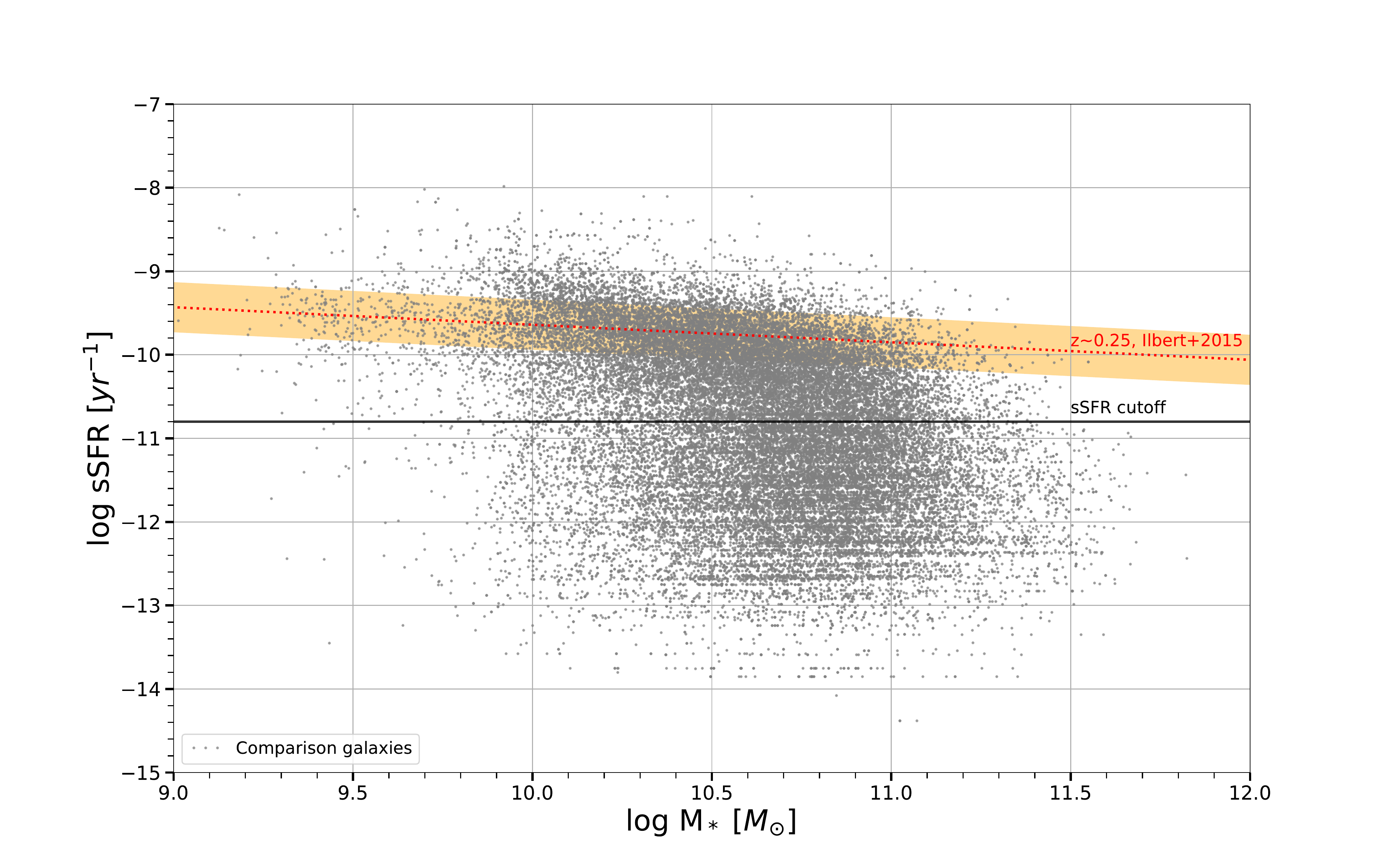}}
   \caption{(a) sSFR distribution of comparison galaxies.
    The sSFR cutoff is shown by a black line.
    (b) SSFR vs. stellar mass for all comparison galaxies.
    We adopt the log(sSFR)$=-$10.8 yr$^{-1}$ as the cutoff to define the SFGs (black solid line).
    For reference, we put the MS relation from literature \citep{Ilbert_2015}, based on the median value of redshift for all star-forming comparison galaxies (red dashed line).
    The shaded area shows the typical 1$\sigma$ dispersion (0.3 dex) for sSFR \citep{Katsianis_2019}.}
    \label{fig:ssfr_cutoff}
\end{figure*}

We further investigate whether quasar neighbors are more similar to neighbors around star-forming galaxies (SFGs) or quiescent galaxies.
We crudely split the sample of matched galaxies into two groups: SFGs and quiescent galaxies (Fig.~\ref{fig:ssfr_cutoff}).
SFGs have a tight correlation between the sSFR versus their stellar mass, known as the main sequence \citep[MS; e.g.][]{Noeske_2007}.
To define the passive galaxies, we adopt a sSFR limit of log(sSFR)$=-$10.8 yr$^{-1}$, guided by the obvious double-peak nature of the sSFR distribution of the galaxies.
We consider all galaxies either star-forming or passive, disregarding (for simplicity) the intermediate (aka 'green valley') objects \citep{Vulcani_2015,Davies_2016}.
For the SFGs, log(sSFR) $\geq-$10.8 yr$^{-1}$.
For passive galaxies, log(sSFR)~\textless $-$10.8 yr$^{-1}$. 
The median redshift of the SFGs is z $\sim$ 0.25.

Then, we identify the neighbors of each subgroup.
We compare the quasar sample of neighbors to the neighbors of each subgroup of comparison galaxies.
All KS tests showed that both populations belong to the same parent distribution.

\subsection{Stellar Mass and Environment}
To check whether the environment effect depends on stellar mass, we divided our sample of quasars (and comparison galaxies) into two mass bins: low mass (M$_\star$ \textless 10$^{10.5}$ M$_\odot$) and high mass (M$_\star$$\geq$ 10$^{10.5}$ M$_\odot$), adopting the massive galaxy definition from \citet[]{Taylor_2011}.
We used the stellar masses from {\fontfamily{qcr}\selectfont MagPhys} DMU.
Also, we note that GAMA provides a more complete census of low-redshift galaxy population than the SDSS, especially for massive galaxies \citep{Taylor_2011}.
We checked that the redshift distribution is similar for both subsamples (low and high mass) between quasar neighbors and comparison galaxy neighbors by running a two-sample KS test.

\begin{figure}
    \centering
    \subfigure[]{\includegraphics[width=0.45\textwidth]{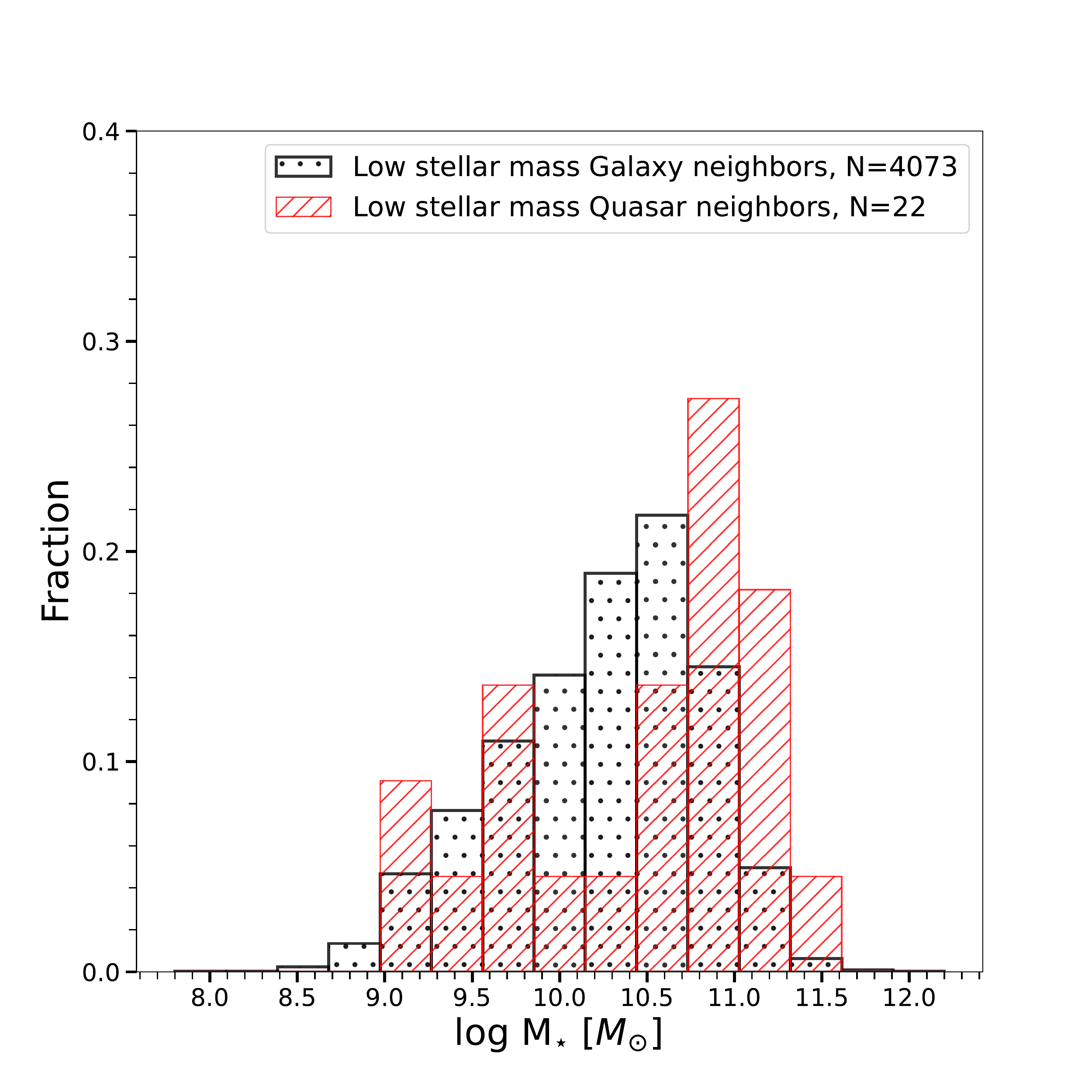}}
    \subfigure[]{\includegraphics[width=0.45\textwidth]{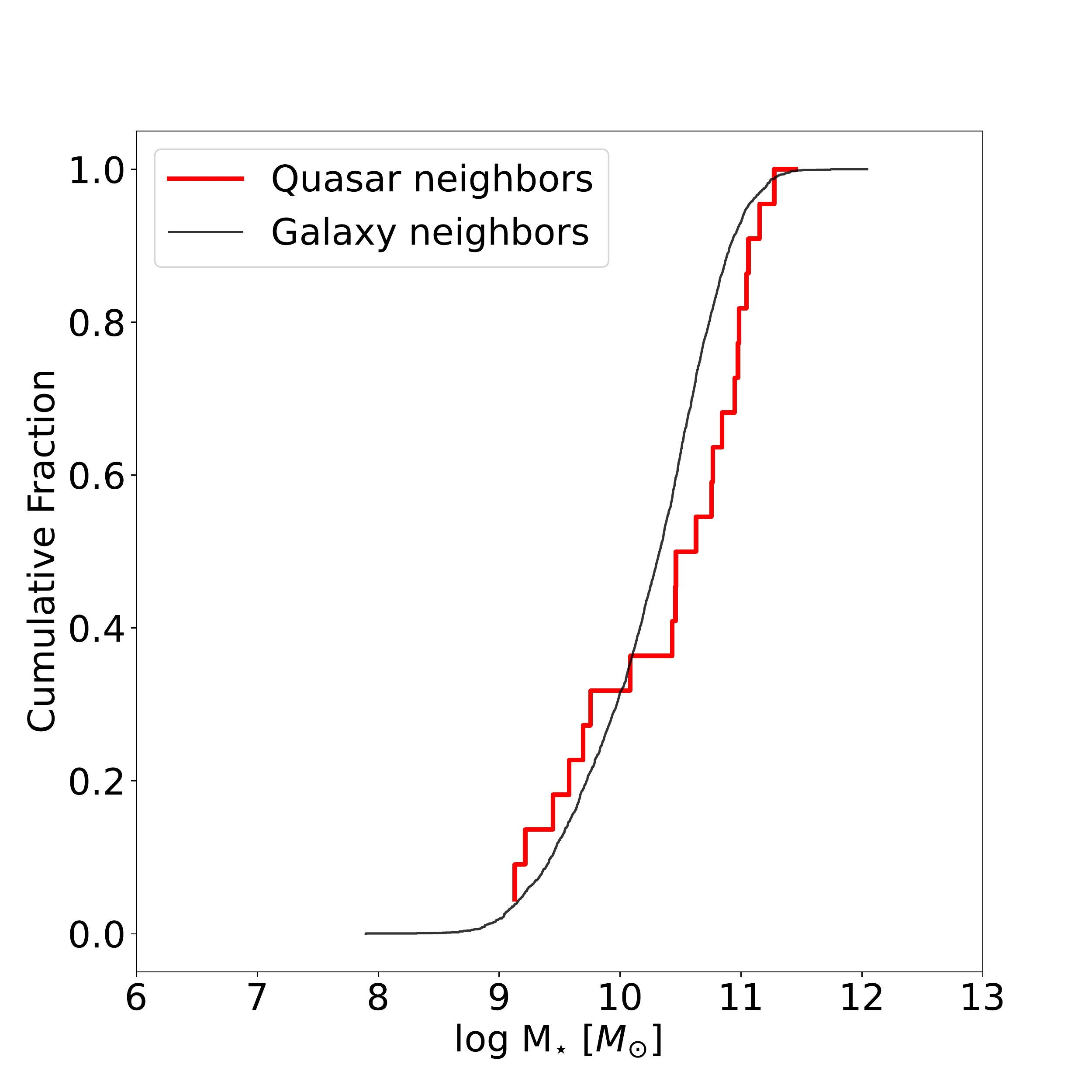}}
    \caption{(a) Stellar mass distribution of neighbors around low-mass quasars and low-mass comparison galaxies.
    (b) KS test result.
    The eCDF, quasar neighbors in red solid line, and comparison galaxy neighbors in black.}
    \label{fig:mass_difference}
\end{figure}

We next compared the neighbors of low mass quasars to the neighbors of low mass comparison galaxies.
Only in one instance, the KS test suggested that there are differences between the two populations - the stellar mass of neighbors (Fig.~\ref{fig:mass_difference}).
In the case of low-mass quasars, their neighbors don't have as many intermediate mass galaxies, when compared to the low-mass comparison galaxies.
This result is based on a small number of low-mass quasar neighbors, but if it is valid, then the absence of intermediate mass companions suggests merger activity \citep{Yue_2019}.
The same analysis was done for the case of high mass subsamples, however the KS test results were in favor of the hypothesis that high mass quasar neighbors and high mass comparison galaxy neighbors come from the same parent distribution.

\section{Discussion} \label{sec:discussion}

We have investigated the properties of neighbors of AGN host galaxies and a comparable sample of inactive galaxies matched in both mass and redshift within the same volume. 
Previous studies showed only a mild enhancement, if any, in the fraction of close companions associated with AGN host galaxies \citep[][]{Serber_2006,Strand_2008,Zhang_2013}; however the neighbors in the above-mentioned studies were selected using photometric data.
Additionally, \citet[]{Strand_2008} and \citet[]{Zhang_2013} lacked a comparison sample of inactive galaxies to isolate any effects due or related to the AGN from those of its environment.
More recently, \citet{Bettoni_2017,Bettoni_2022} and \citetalias{Stone_2021} used long-slit spectroscopy to measure redshifts (and SF properties from emission lines) for galaxies projected near low-redshift ($z\lesssim0.5$) AGN.

For the low-luminosity nearby quasars (z\textless 0.5, log(M$_\star$/M$_\odot$)\textless 11.5), based on spectroscopic redshifts, our findings are consistent with recent results from \citet{Karhunen_2014,Bettoni_2015}, and \citetalias{Wethers_2022} who showed that AGN host galaxies occupy environments similar to those of a comparison sample matched in stellar mass and redshift, implying that the local galaxy environment provides little contribution to the AGN activity. 
The situation may, however, be different for brighter AGN and higher redshift galaxies \citep[e.g.][]{Yue_2019}. 
Moreover, rich clusters are absent in GAMA regions, although the typical environment of galaxies (including quasar host galaxies as shown by \citetalias{Wethers_2022}) resembles that of a moderately rich group: AGN activity has a higher impact on the physical properties of neighbors in less massive objects.

One caveat is that our sample consists of unobscured quasars. 
Within the unification theory framework \citep{Antonucci_1993,Urry_1995}, the unobscured quasars are expected to have the same properties as obscured quasars (as they evolve into each other). 
However, since some results suggest that obscured quasars are distinct objects, it is possible that their neighborhoods also differ from those of the unobscured quasars considered in this study.

The nature of neighboring galaxies may offer clues as to the nature of the AGN host galaxy (whose properties are often difficult to measure because of contamination from the bright nucleus) and to the effect of the AGN on its environment (e.g., where jets interact with the surrounding intergalactic medium, especially in denser environments, as in \citealt{Martin_2021}).
The quasar hosts in our sample have various morphologies and SFR, as can be gleaned from few examples in Fig.~\ref{fig:quasar_hosts_images} and Fig.~\ref{fig:compgal_images}.

\begin{figure}[htb!]
    \centering
    \subfigure[]{\includegraphics[width=0.47\textwidth]{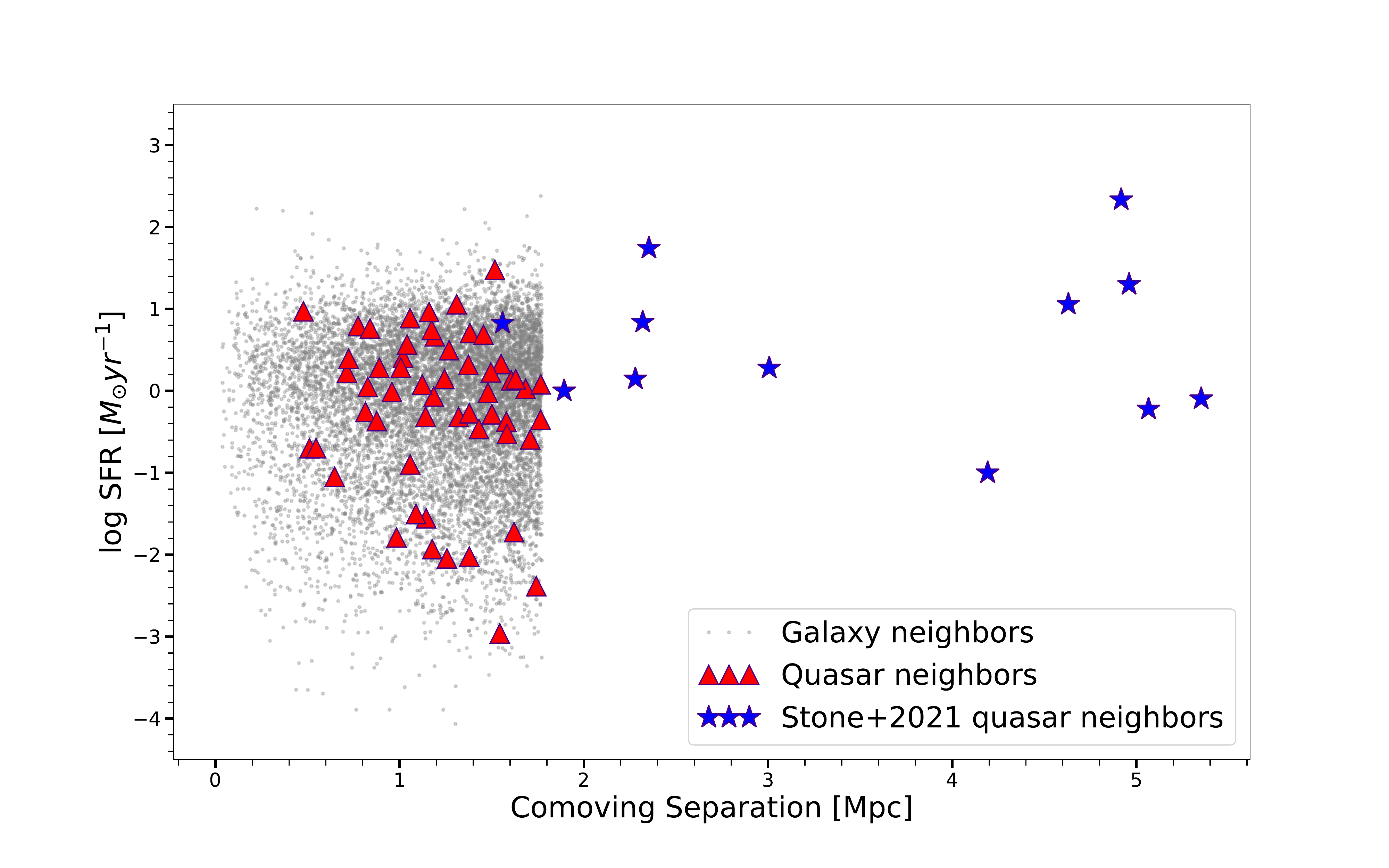}}
    \subfigure[]{\includegraphics[width=0.47\textwidth]{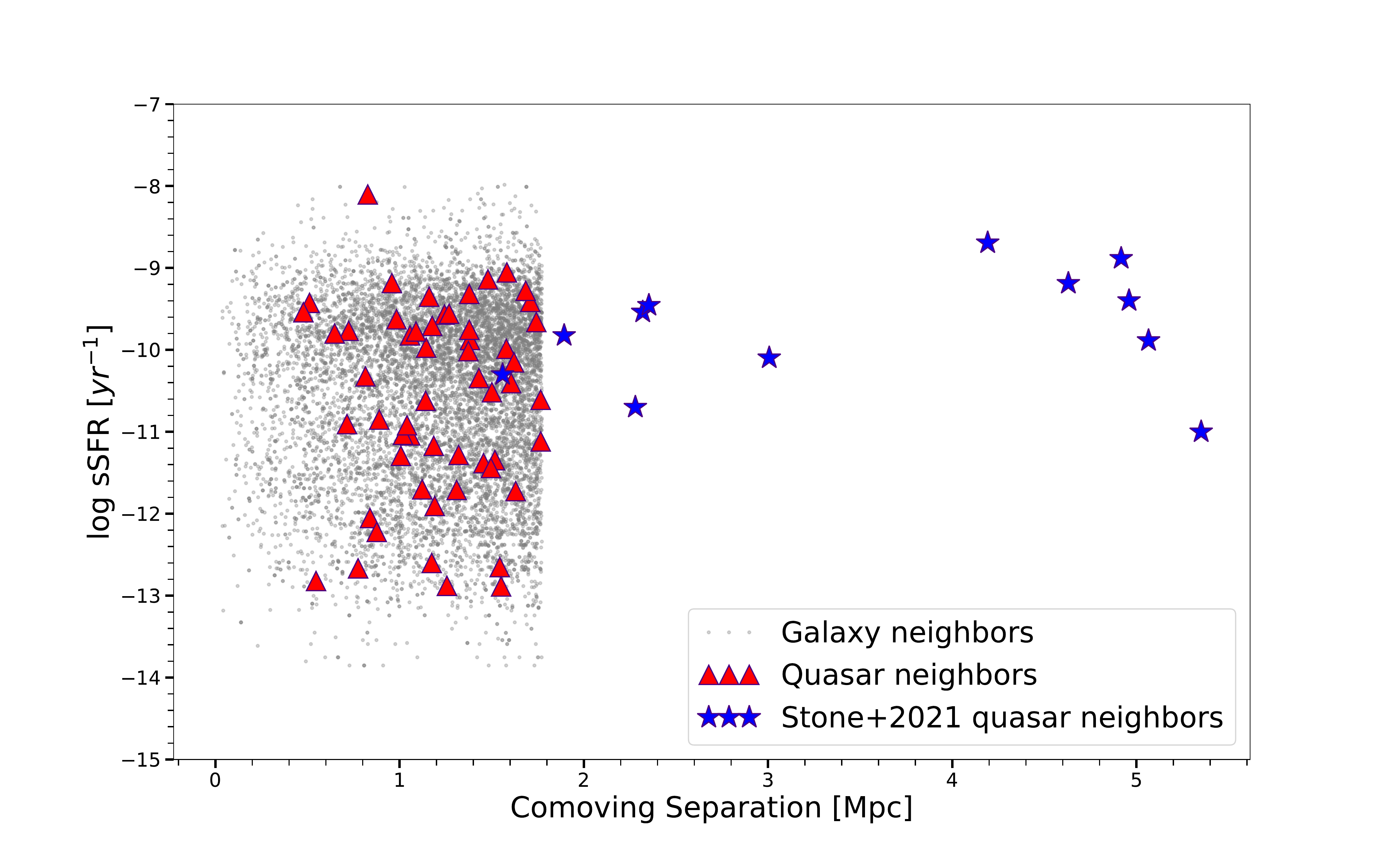}}\\
    \subfigure[]{\includegraphics[width=0.47\textwidth]{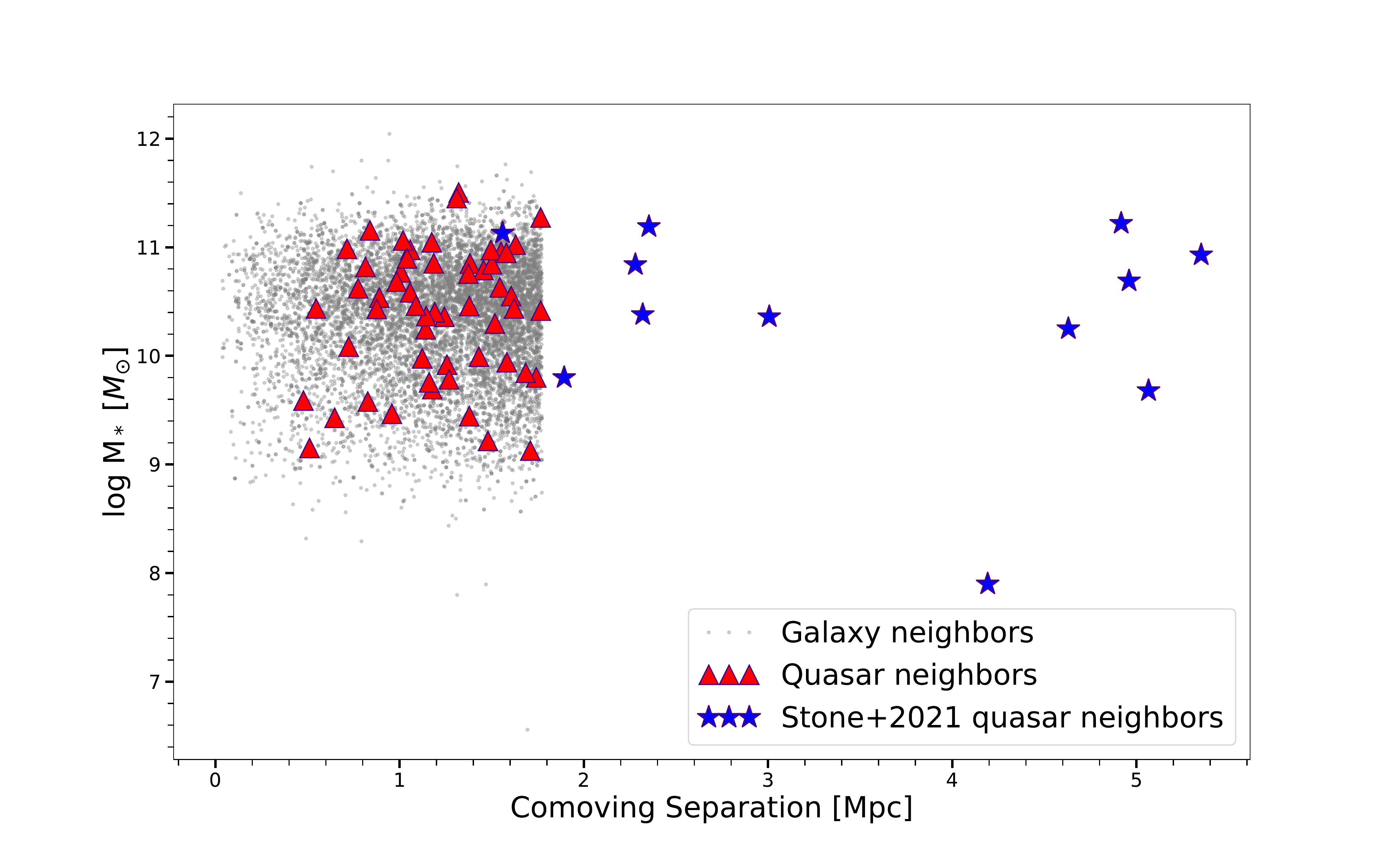}}
\caption{Correlation of (a) SFR vs. comoving separation of the neighboring galaxy from the quasar (or comparison galaxy), (b) sSFR vs. comoving separation and (c) stellar mass vs. comoving separation.
The comoving separation here is the distance from the central object (quasar or comparison galaxy).
No differences are observed in these correlation plots between the neighbors of quasars and matched inactive galaxies.}
\label{fig:correlations1}
\end{figure}

We find that there is no evidence of any significant difference between the neighbors of AGN host galaxies and those of the mass and redshift matched sample of inactive galaxies, in terms of morphology, current SFRs, SFH indicators, and age. 
This suggests that AGN host galaxies will also have very similar properties to those of inactive galaxies (in agreement with studies of quasar host galaxies at intermediate redshifts by \citealt{Alam_2021}); similarly that the AGN activity also has little influence on the properties of its neighbors, although this may occur in environments where the intergalactic medium is denser \citep{Dressler_1980}. 
Fig.~\ref{fig:correlations1} shows this particularly well, where we find that the sSFR of the neighbors of the AGN hosts does not depend on distance from the galaxy and follows the same distribution as for neighbors of inactive galaxies. 

This is in agreement with earlier results from \citet{Villarroel_2012}, although \citet{Coldwell_2003,Coldwell_2006} claimed an excess of SFGs within 1 Mpc of AGN.
Neither do we see evidence that AGN hosts have more massive neighbors, unlike the trend claimed by \citet{Yue_2019}, albeit at a higher redshift.

\begin{figure*}[htb!]
    \centering
    \subfigure[]{\includegraphics[width=0.44\textwidth]{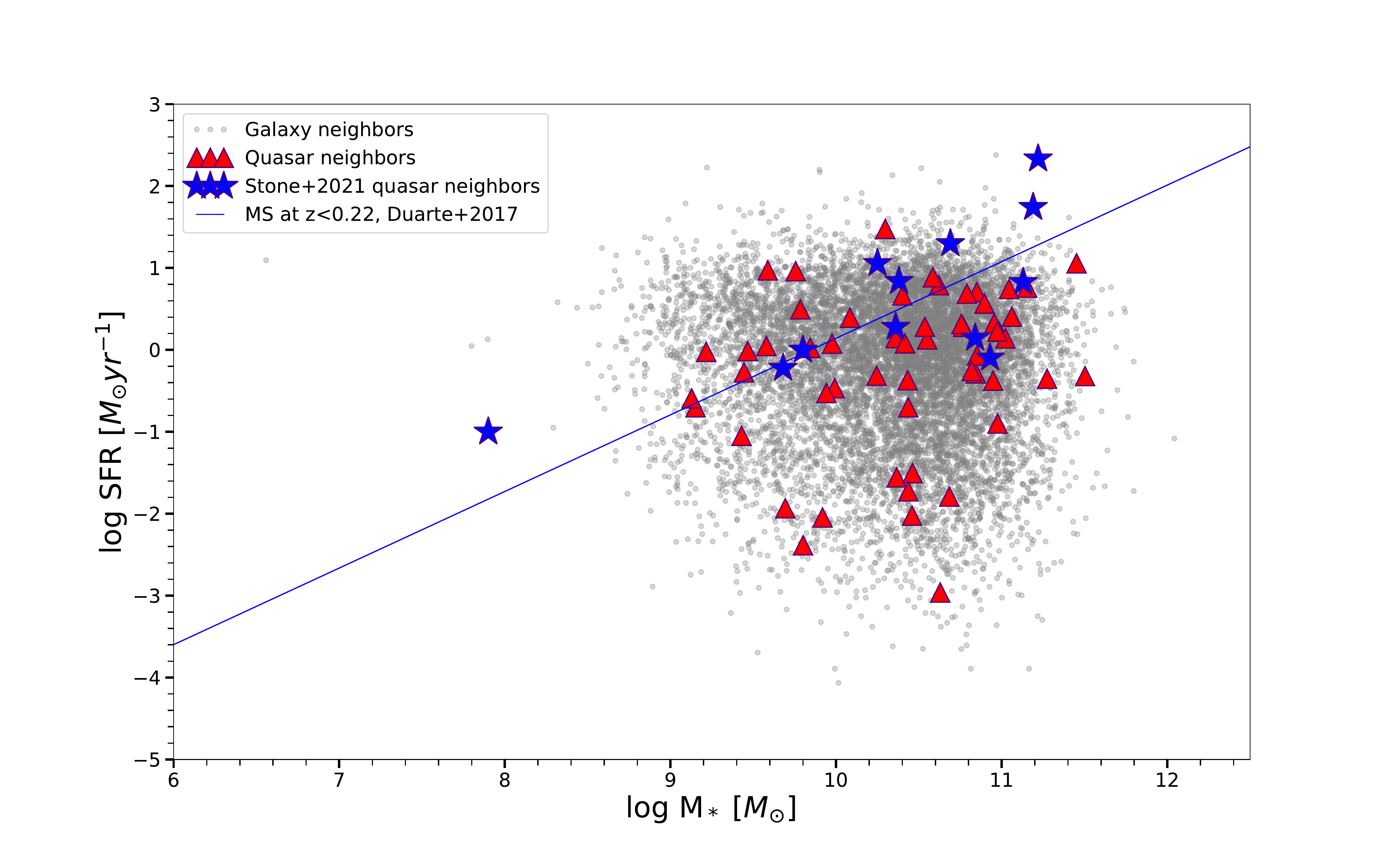}}
    \subfigure[]{\includegraphics[width=0.44\textwidth]{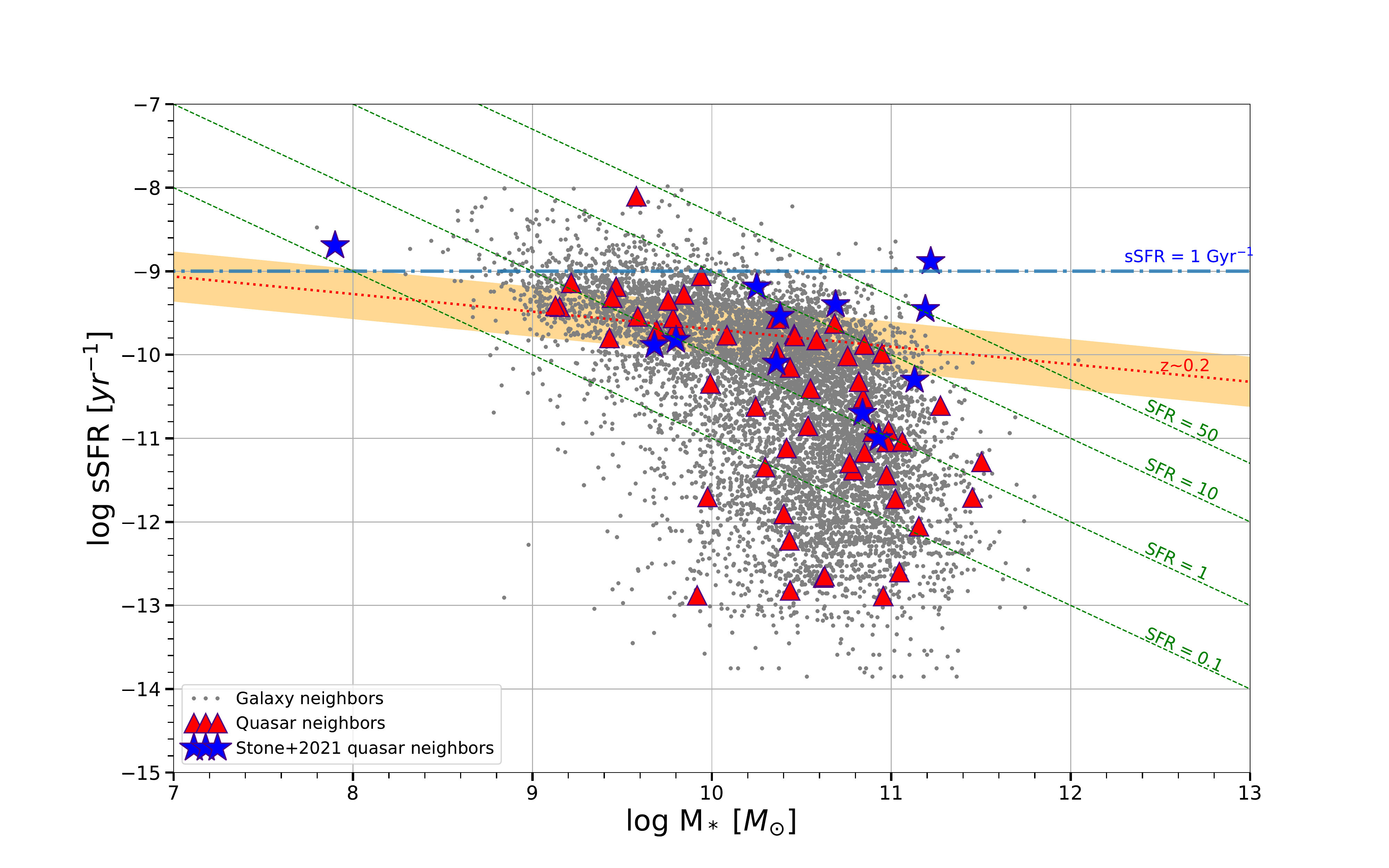}}
    \caption{Correlations of mass with (a) SFR and (b) sSFR.
Constant SFR (M$_{\odot}$ yr$^{-1}$) lines are green (dashed). 
The sSFR constant line is blue (dash-dotted).
The red dotted line is the MS relation from literature \citep{Ilbert_2015}, using the median redshift of all comparison galaxy neighbors (z$\sim$0.3). 
The shaded region around it is the typical 1$\sigma$ dispersion \citep[0.3 dex;][]{Katsianis_2019}.
}
\label{fig:correlations2}
\end{figure*}

We also explore the correlations of stellar mass and SFR for the neighbors (Fig.~\ref{fig:correlations2} a).\@
The SFGs form a distinct sequence (MS) in the SFR (sSFR) vs. M$_\star$ plane at a given redshift \citep{Noeske_2007,Kennicutt_2012}.
In our sample of low-redshift quasar and galaxy neighbors, the sSFR vs. stellar mass plot (Fig.~\ref{fig:correlations2} b) shows that galaxies with lower stellar mass form stars at a higher rate as compared to more massive galaxies, as expected \citep[e.g.][]{Schreiber_2016,Abdurrouf_2017}.
We plot the MS relation from literature for low-redshift SFGs for reference in the log(SFR) vs. logM$_{\star}$ plane \citep{Duarte_2017} and in the log(sSFR) vs. logM$_{\star}$ plane \citep{Ilbert_2015}.
The shaded region in Fig.~\ref{fig:correlations2} b) shows the 1$\sigma$ dispersion of sSFR.
Most of the galaxy neighbors and quasar neighbors conform to the definition of SFG to some degree.
Conversely, none of the neighboring galaxies in quasar fields have SFR larger than 50 M$_{\odot}$ yr$^{-1}$, which is against the scenario where mergers boost very high levels of SFR in AGN interactions.

Simulations in the scenarios where quasar activity is induced by mergers indicate starburst-like activity in the companion galaxies as well as the host. 
However, in our study, we do not observe any enhancement of SFR in the quasar companions relative to the neighbors of comparison galaxies.
Further, we don't see any evidence that the neighbors closer to the quasar have higher SFR.
While it is possible that the SF activity peaks on a shorter timescale, given our large number of quasars and their neighbors, our results are robust.
Even when quasar companion samples are compared against only companions of the star-forming comparison galaxies or of only quiescent comparison galaxies, the SFR-related properties do not show any statistically significant difference.

These results suggest that at least at low redshift AGN activity is not triggered by mergers and does not strongly depend on the environment.
If secular processes are responsible, AGN are therefore a random occurrence in the lifetime of galaxies, and the correlations between galaxy properties and SMBH masses may occur via a random walk process \citep{Jahnke_2011}.
This is consistent with the recent simulations by \citet{Draper_2012} showing that at lower redshifts processes other than mergers and interactions are the dominant AGN triggering mechanism. 
On the other hand, our results disagree with the conclusions of \citet{Gao_2020}, where merger fractions are higher in AGN than in non-AGN, although this may be explained by their inclusion of AGN selected in the mid-infrared, where merger rates seem to be consistently higher.
In contrast, \citet{Silva_2021} found that AGN fraction is similar in mergers and non-mergers.
The existence of a merger-AGN connection remains a controversial issue.

\section{Conclusions} \label{sec:conclusions}

We have studied the properties and environments of galaxies neighboring 205 low-redshift Type I (unobscured) AGN and a comparable sample of inactive galaxies, drawn from the GAMA survey in order to assess whether AGN hosts lie in different close environments and how they affect their neighbors. 
We found no evidence of any significant difference in the properties (morphology, colors, SFR, SFH, stellar mass, dust mass, age, and metallicity) of galaxies surrounding AGN hosts compared to those of inactive galaxies matched in mass and redshift and within the same volume. 
We explored the dependence of the environment effect on stellar mass and SF.
Overall, our results suggest that the properties of galaxies around quasars are not influenced by the quasar activity. 
We conclude that AGN activity is more likely to be triggered by internal or secular processes, with major mergers playing only a minor role.

\begin{acknowledgments}

MBS acknowledges useful discussion on galaxy clustering with Maret Einasto.
MBS acknowledges the Finnish Cultural Foundation grant number 00220968 and Finnish Centre for Astronomy with ESO grant.
MBS and JK acknowledge financial support from the Academy of Finland, grant 311438.

GAMA is a joint European-Australasian project based around a spectroscopic campaign using the Anglo-Australian Telescope.
The GAMA input catalogue is based on data taken from the Sloan Digital Sky Survey and the UKIRT Infrared Deep Sky Survey.
Complementary imaging of the GAMA regions is being obtained by a number of independent survey programmes including GALEX MIS, VST KiDS, VISTA VIKING, WISE, Herschel-ATLAS, GMRT and ASKAP providing UV to radio coverage.
GAMA is funded by the STFC (UK), the ARC (Australia), the AAO, and the participating institutions.
The GAMA website is http://www.gama-survey.org/ .

This research has made use of the NASA/IPAC Extragalactic Database (NED) which is operated by the Jet Propulsion Laboratory, California Institute of Technology, under contract with the National Aeronautics and Space Administration.
This research made use of the NASA's Astrophysics Data System Bibliographic Services, Astrobetter blog and wiki, and cosmology calculator by \citet{Wright_2006}.
This research has made use of the SIMBAD database, operated at CDS, Strasbourg, France.
This research made use of code from \citet{Stone_2017}.
This research has made use of the VizieR catalogue access tool, CDS, Strasbourg, France (DOI : 10.26093/cds/vizier).
The original description of the VizieR service was published in 2000, A\&AS 143, 23
 
\end{acknowledgments}

\facilities{AAT}

\software{Astropy \citep{Astropy_2013,Astropy_2018}, 
NUMPY \citep{Harris_2020}, 
SCIPY \citep{Virtanen_2020}, 
MATPLOTLIB \citep{Hunter_2007}.
}

\bibliography{references}{}
\bibliographystyle{aasjournal}
\clearpage
\appendix

\section{DISTRIBUTIONS OF PROPERTIES}
\label{sec:appendix}

From here we show the distribution plots for the properties considered in this work, as mentioned in the Table~\ref{table:properties}.
We used the KS test to compare the neighbors of quasars with the neighbors of matched inactive galaxies, finding no statistically significant differences in each case

\begin{figure}
    \centering
    \includegraphics[width=0.5\textwidth]{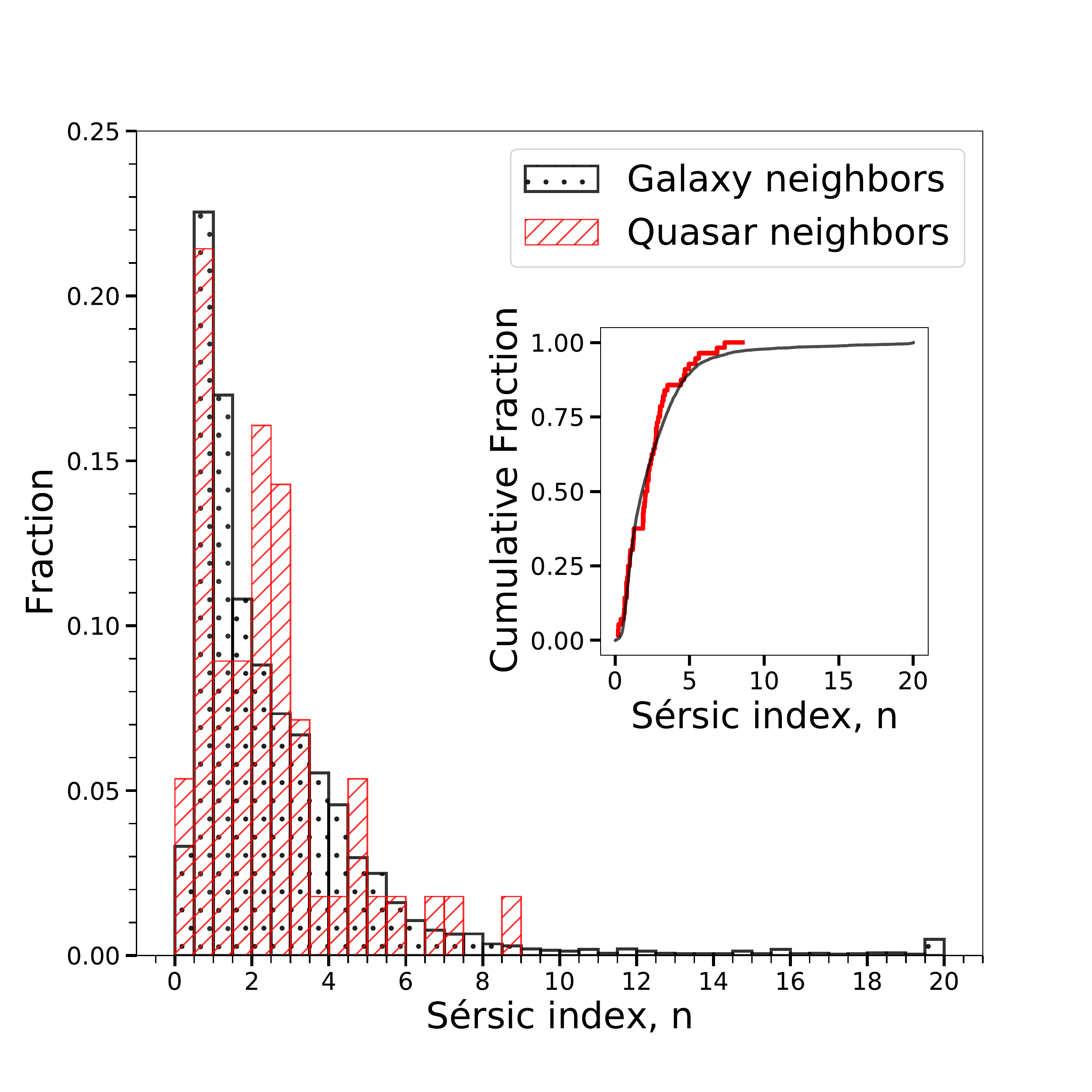}
    \caption{S\'ersic index (quasar neighbors in red hatched bars, comparison galaxy neighbors in black dotted bars).
    The inset shows the eCDF, quasar neighbors in red solid line, and comparison galaxy neighbors in black.
    Median error for Sérsic index values from GAMA is $\sim$ 0.14 both for quasar neighbors and for comparison galaxy neighbors.
    }
    \label{fig:sersic_index_results}
\end{figure}

\begin{figure}[ht]
    \centering
    \subfigure[]{\includegraphics[width=0.47\textwidth]{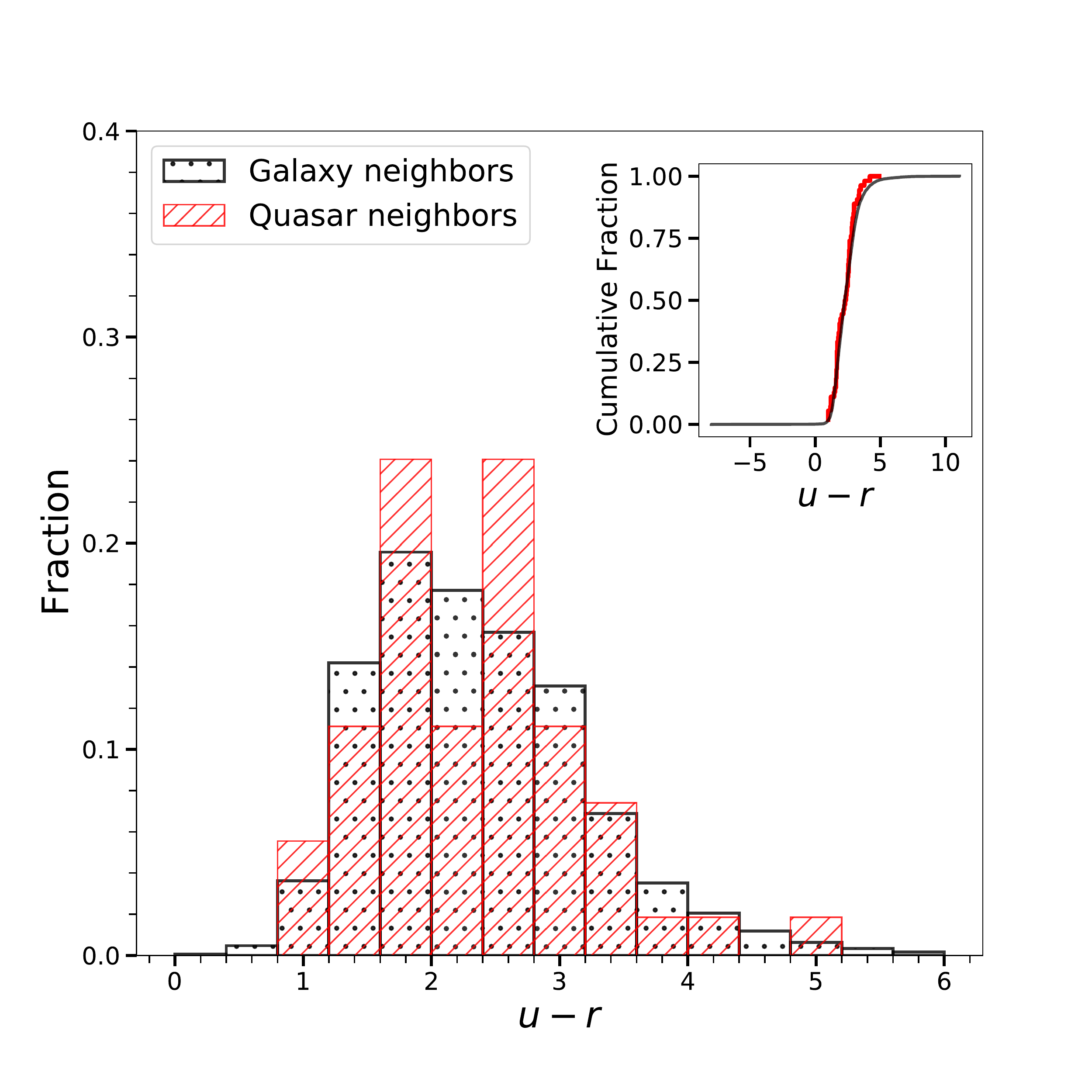}}
    \subfigure[]{\includegraphics[width=0.47\textwidth]{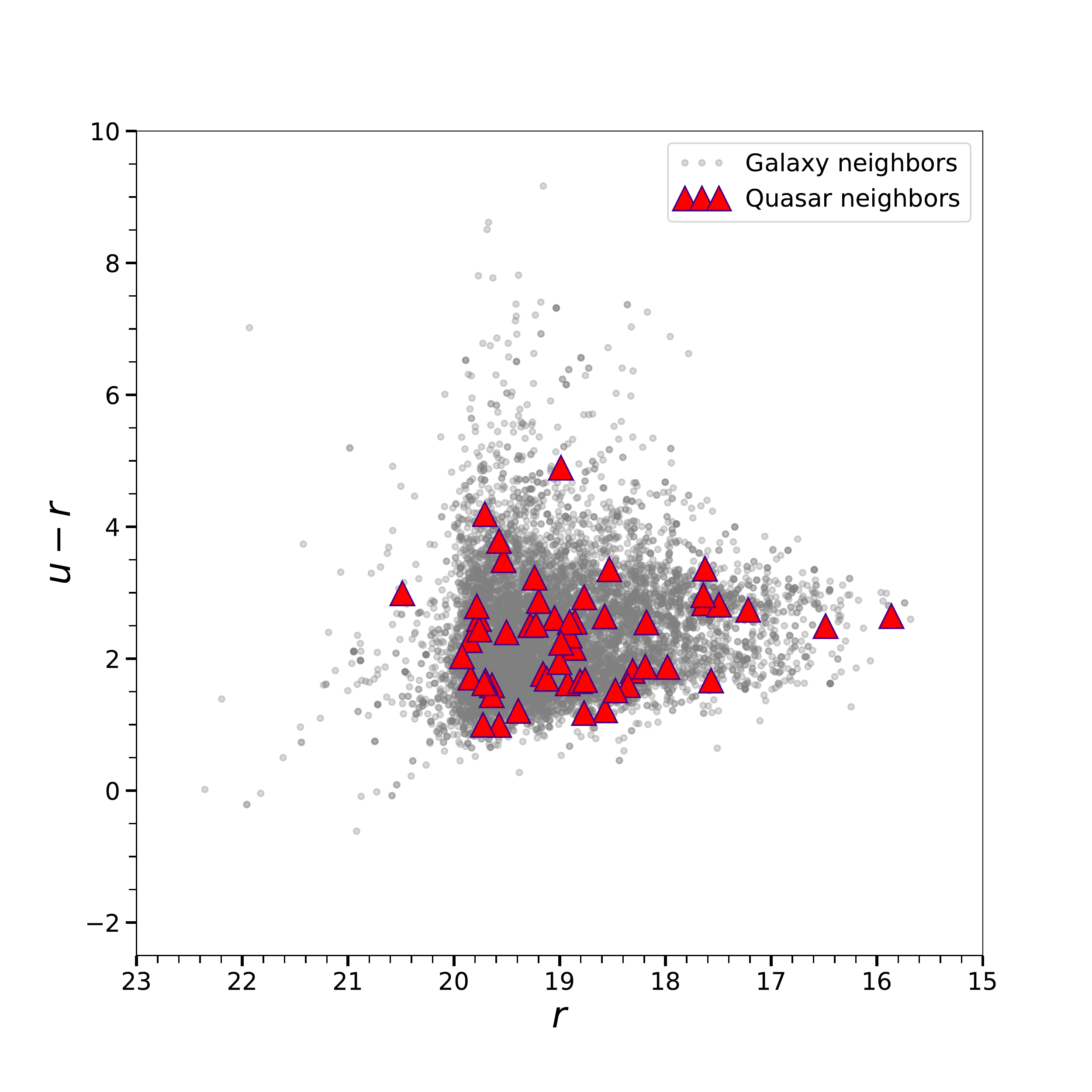}}
\caption{(a) SDSS $u-r$ color (quasar neighbors in red hatched bars, comparison galaxy neighbors in black dotted bars).
The inset shows the eCDF, quasar neighbors in red solid line, and comparison galaxy neighbors in black.
(b) Color-magnitude plot.
Median error for SDSS $u$-$r$ color from GAMA is $\sim$ 0.34 for quasar neighbors and $\sim$ 0.46 for comparison galaxy neighbors.
Median error for SDSS $r$ is $\sim$0.03 for both quasar neighbors and for comparison galaxy neighbors.
To propagate the error, standard uncertainty propagation rules were used.
}
\label{fig:color_u_r}
\end{figure}

\begin{figure}
    \centering
    \subfigure[]{\includegraphics[width=0.49\textwidth]{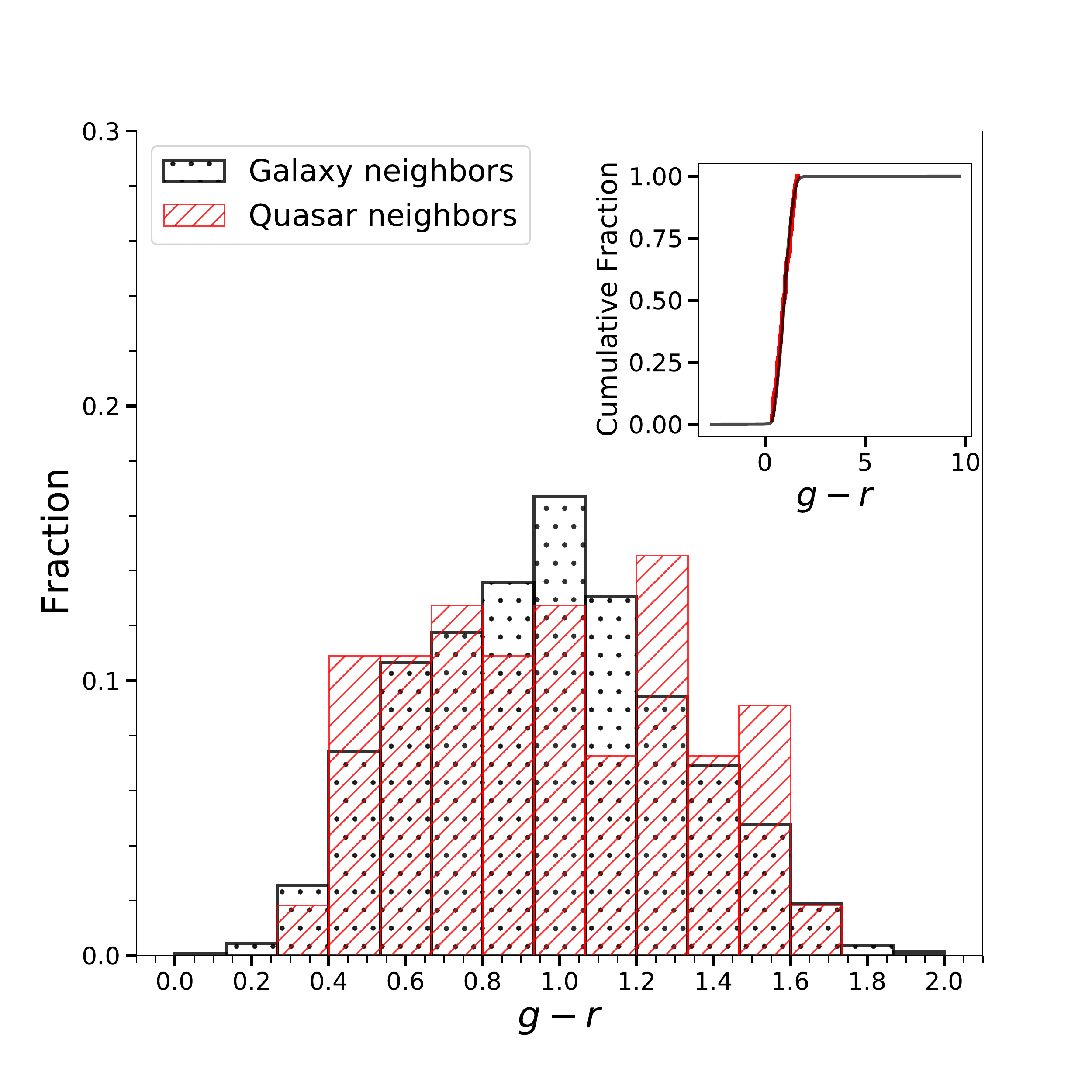}}
    \subfigure[]{\includegraphics[width=0.49\textwidth]{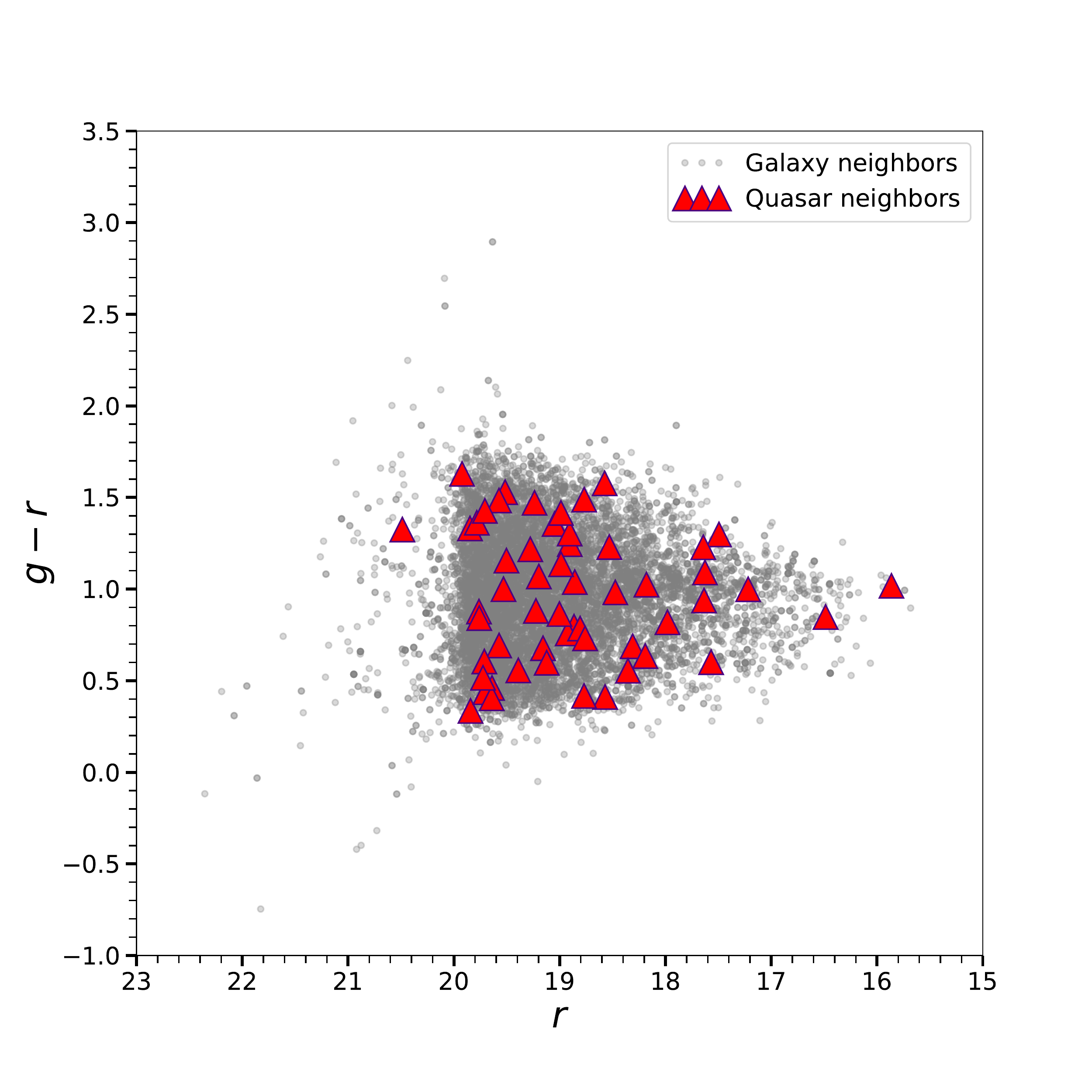}}
\caption{(a) SDSS $g-r$ color (quasar neighbors in red hatched bars, comparison galaxy neighbors in black dotted bars).
The inset shows the eCDF, quasar neighbors in red solid line, and comparison galaxy neighbors in black.
(b) Color-magnitude plot.
Median error for SDSS $g$-$r$ color from GAMA is $\sim$ 0.05 for quasar neighbors and $\sim$ 0.06 for comparison galaxy neighbors.
Median error for SDSS $r$ is $\sim$0.03 for both quasar neighbors and for comparison galaxy neighbors.
To propagate the error, standard uncertainty propagation rules were used.
}
\label{fig:color_g_r}
\end{figure}

\begin{figure}
    \centering
    \subfigure[]{\includegraphics[width=0.47\textwidth]{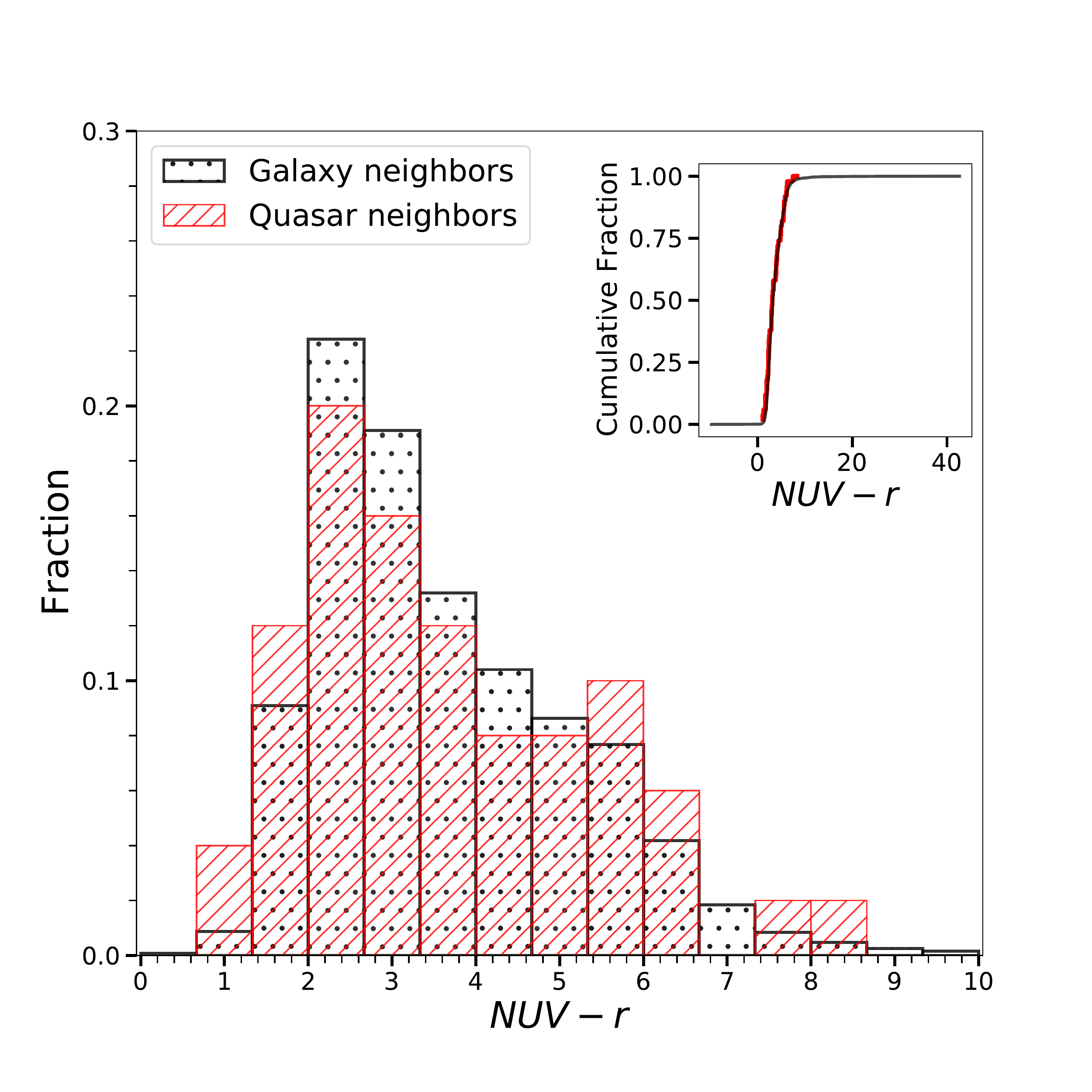}}
    \subfigure[]{\includegraphics[width=0.47\textwidth]{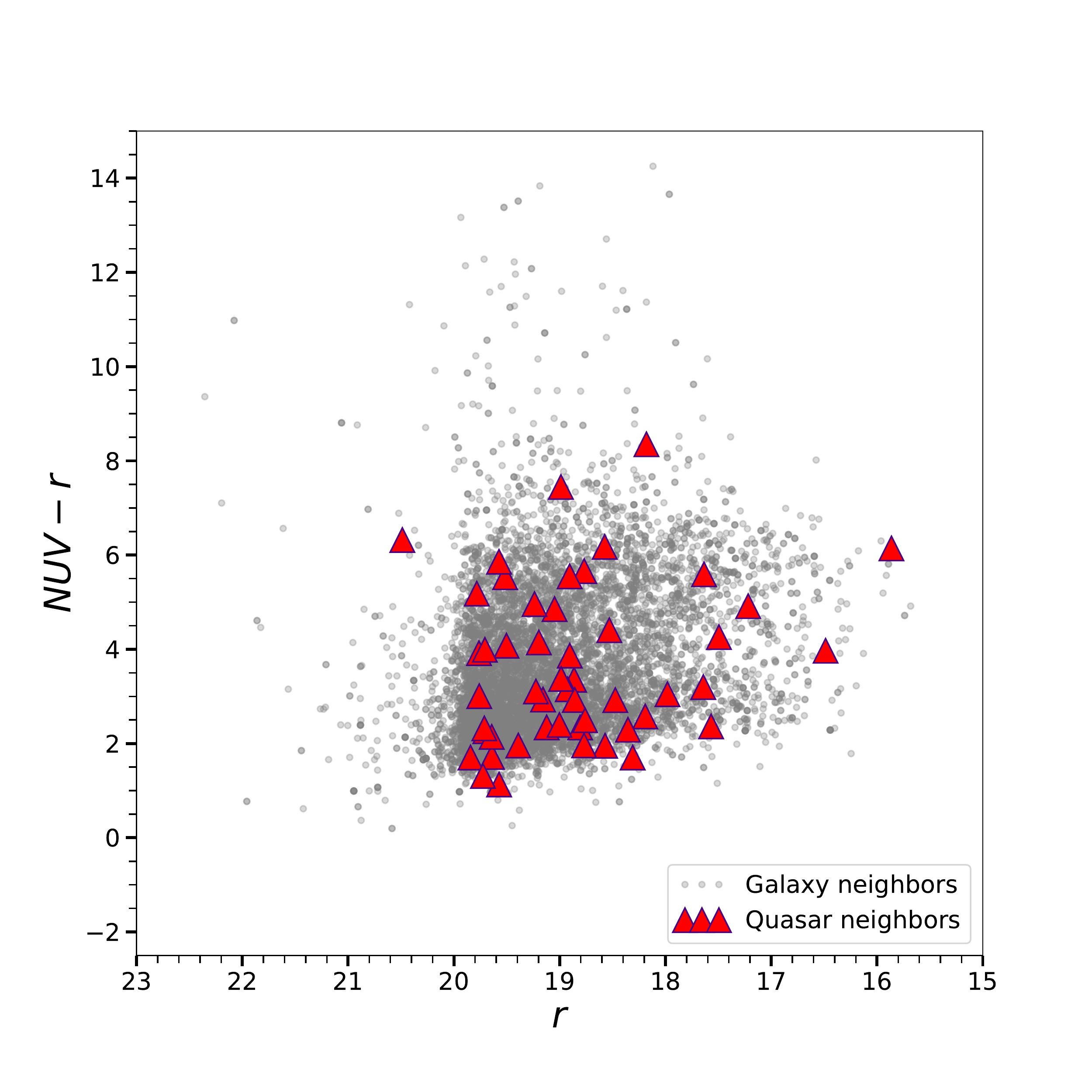}}
\caption{(a) \textit{GALEX} $NUV$ - SDSS $r$ color (quasar neighbors in red hatched bars, comparison galaxy neighbors in black dotted bars).
The inset shows the eCDF, quasar neighbors in red solid line, and comparison galaxy neighbors in black.
(b) Color-magnitude plot.
Median error for $NUV$ - SDSS $r$ color from GAMA is $\sim$ 2.1 for quasar neighbors and $\sim$ 2.4 for comparison galaxy neighbors.
Median error for $r$ is $\sim$0.03 for both quasar neighbors and for comparison galaxy neighbors.
To propagate the error, standard uncertainty propagation rules were used.
}
\label{fig:color_nuv_r}
\end{figure}

\begin{figure}
\centering
    \includegraphics[width=0.47\textwidth]{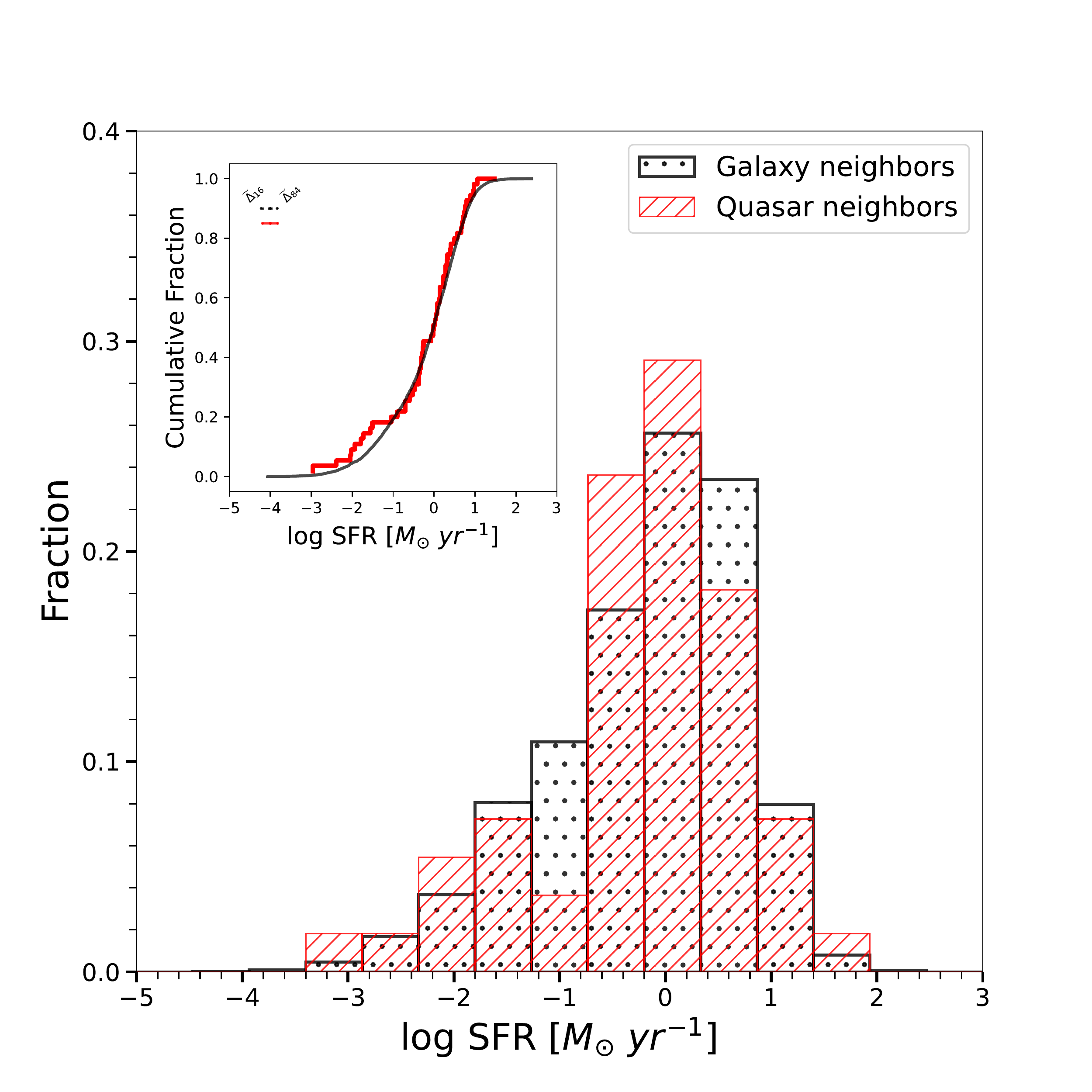}
    \includegraphics[width=0.47\textwidth]{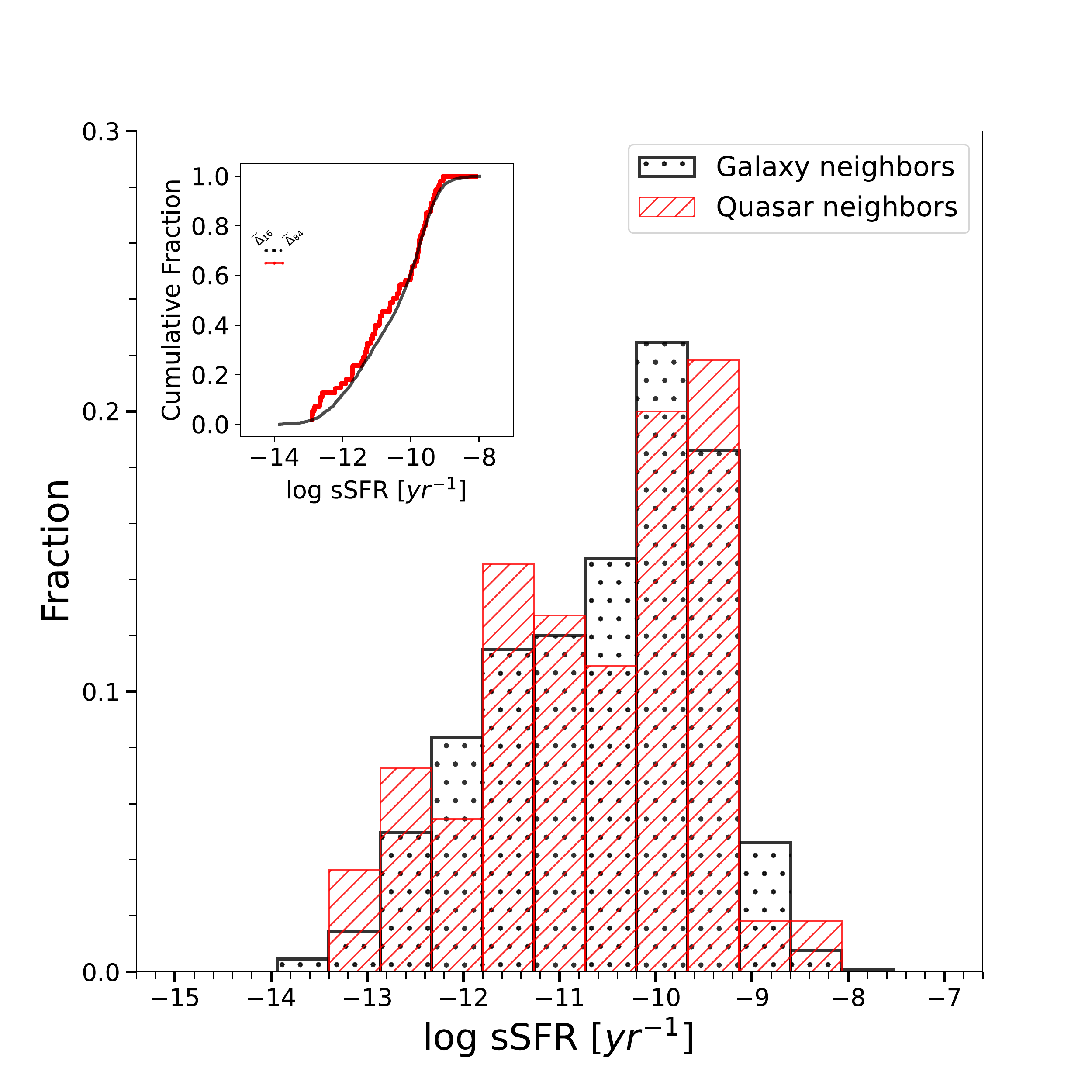}
\caption{(a) SFR. 
(b) sSFR (quasar neighbors in red hatched bars, comparison galaxy neighbors in black dotted bars).
The inset shows the eCDF, quasar neighbors in red solid line, and comparison galaxy neighbors in black.
The typical uncertainty is represented by the median values of 16th and 84th percentiles.}
\label{fig:SFR_results}
\end{figure}

\begin{figure}
    \centering
    \subfigure[]{\includegraphics[width=0.49\textwidth]{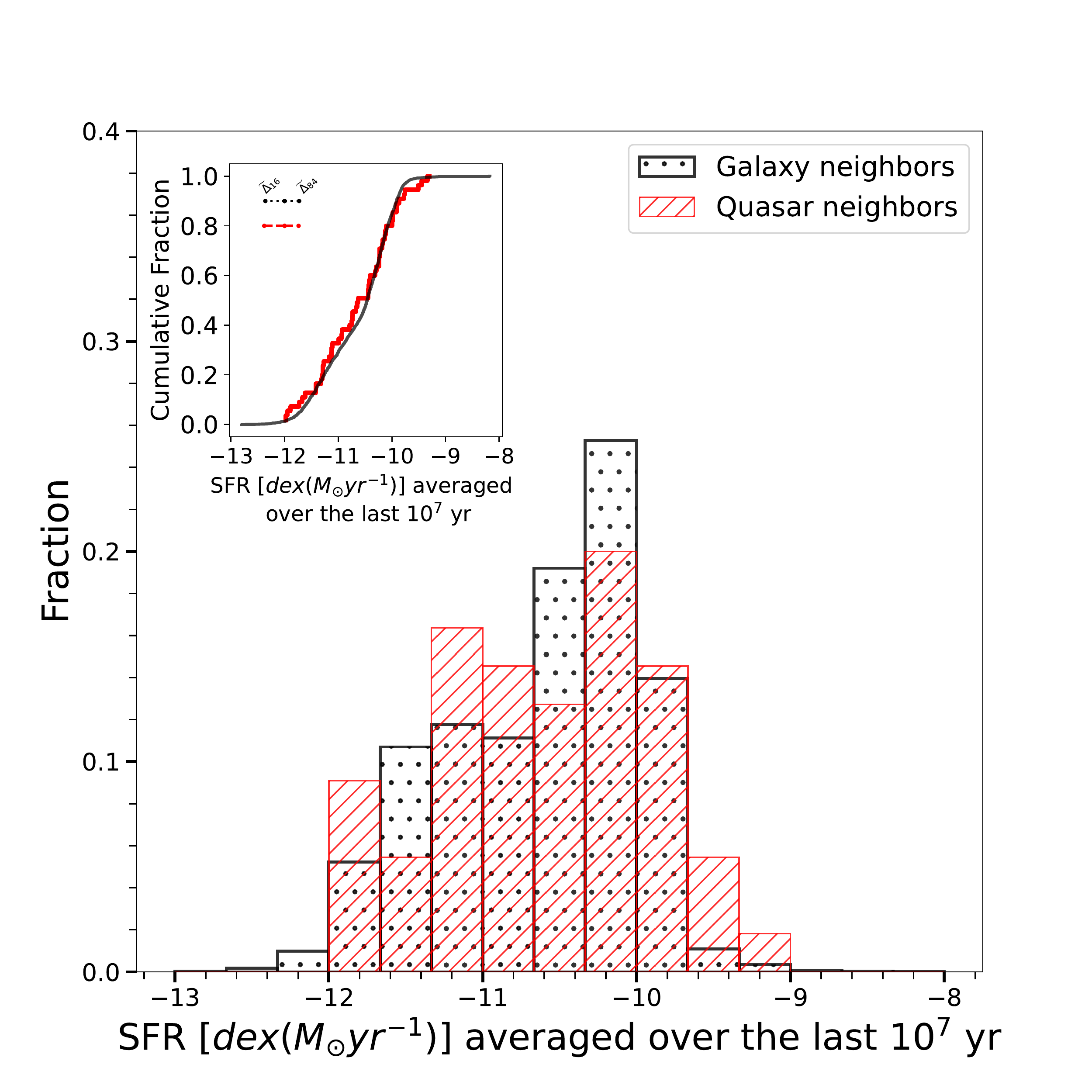}}
    \subfigure[]{\includegraphics[width=0.49\textwidth]{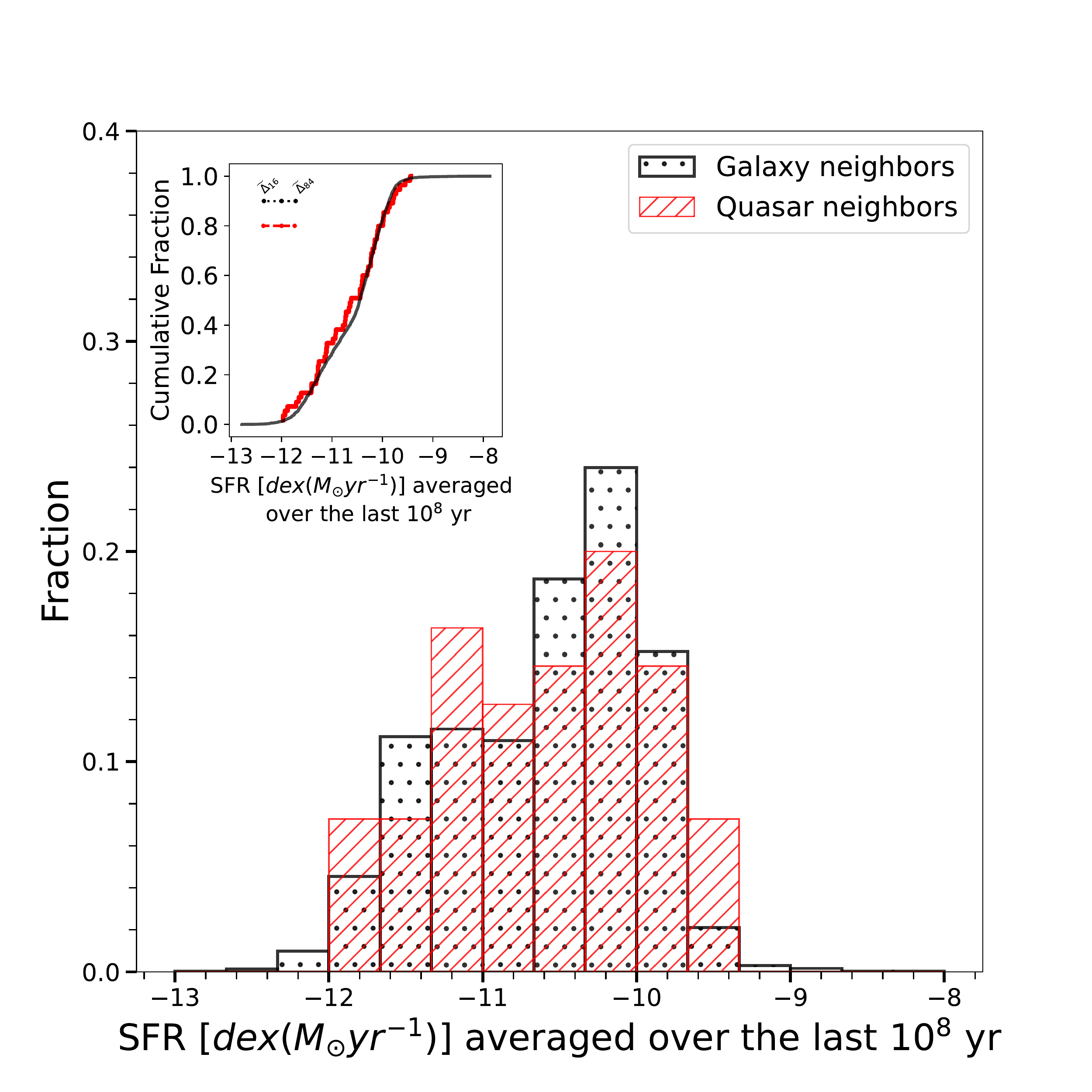}}\\
    \subfigure[]{\includegraphics[width=0.49\textwidth]{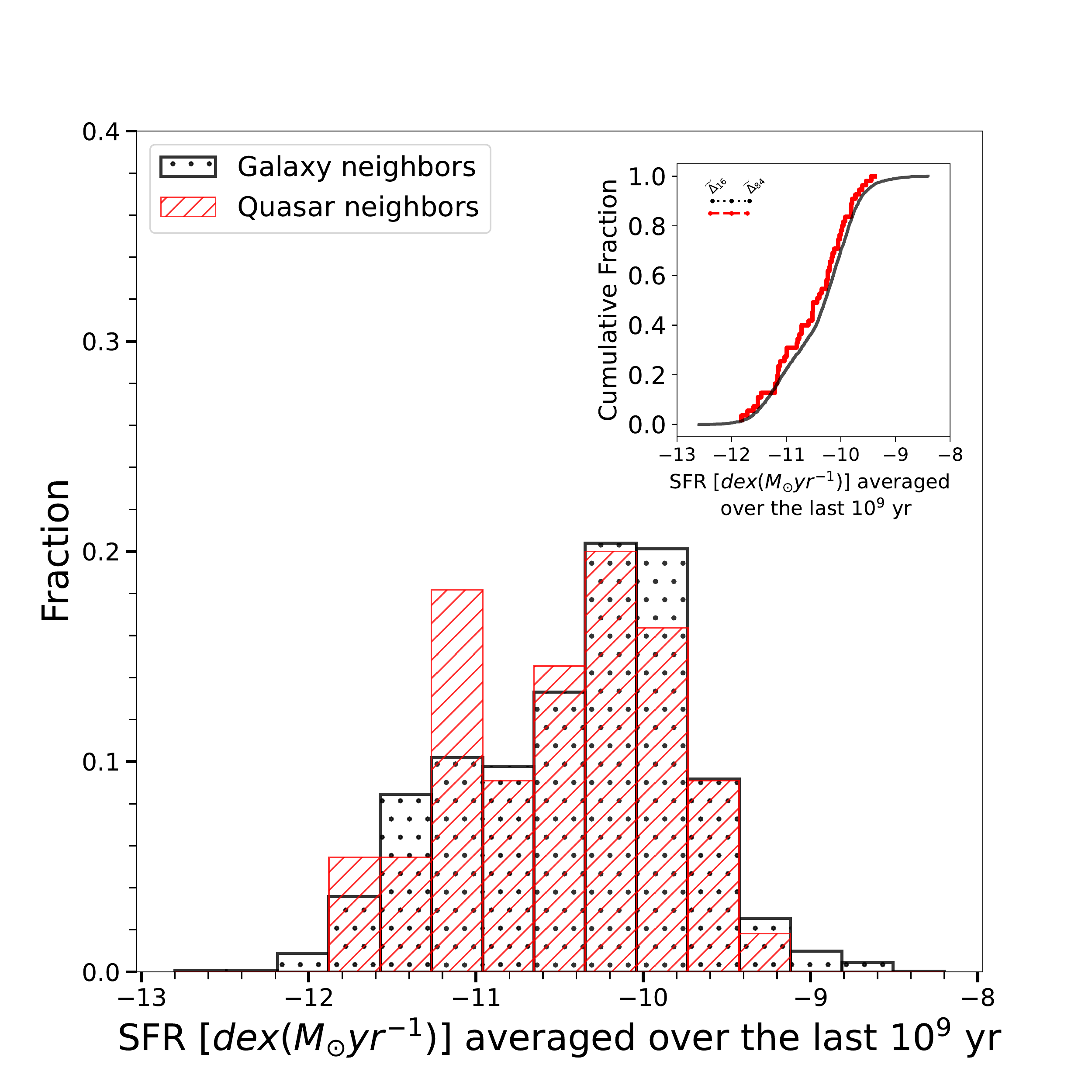}}
    \subfigure[]{\includegraphics[width=0.49\textwidth]{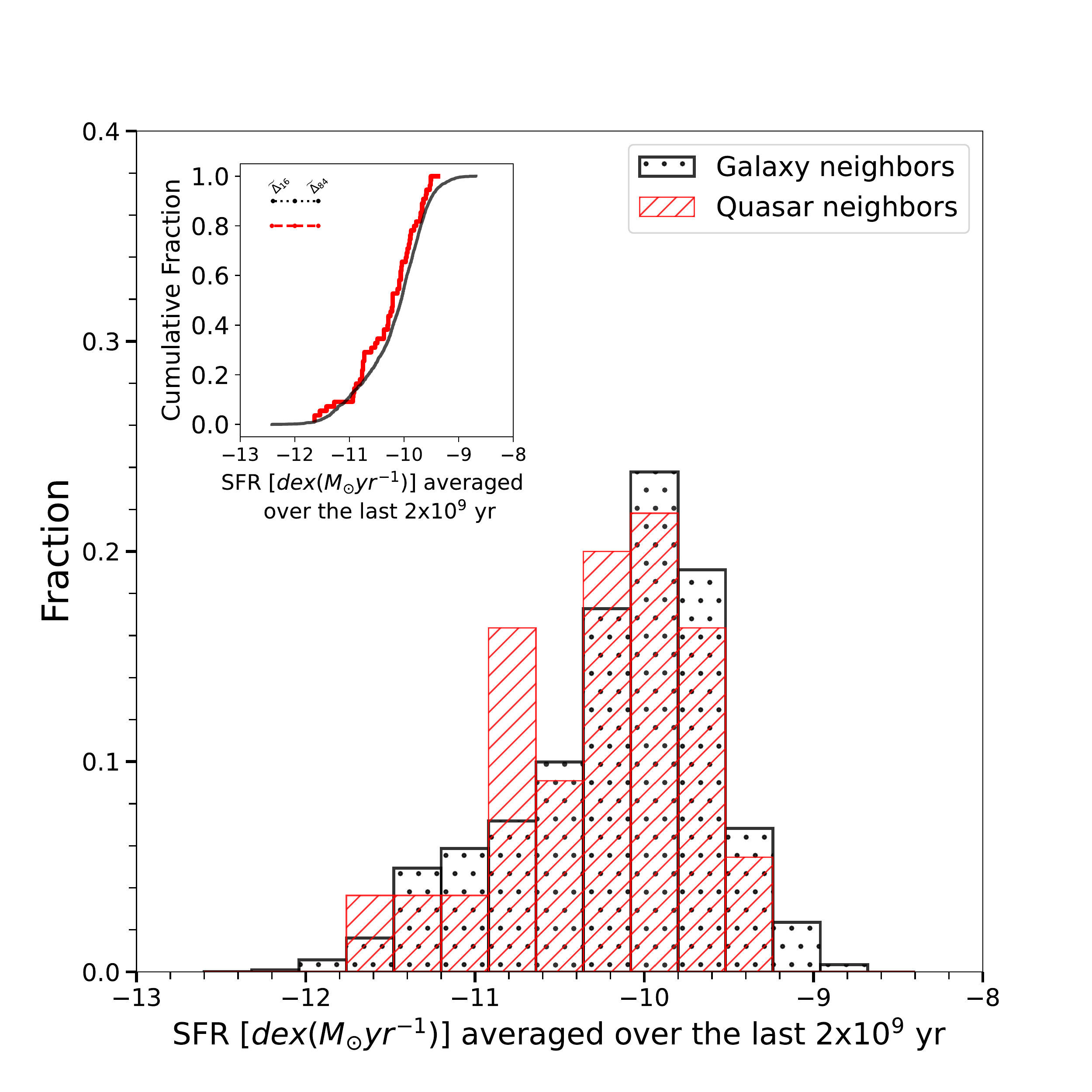}}
\caption{Median SFR averaged over the past $10^7$ yr, $10^8$ yr, $10^9$ yr, $2 \times 10^9$ yr (quasar neighbors in red hatched bars, comparison galaxy neighbors in black dotted bars).
The inset shows the eCDF, quasar neighbors in red solid line, and comparison galaxy neighbors in black.
The typical uncertainty is represented by the median values of 16th and 84th percentiles.}
\label{fig:sfr17_results}
\end{figure}

\begin{figure}
\centering
    \includegraphics[width=0.49\textwidth]{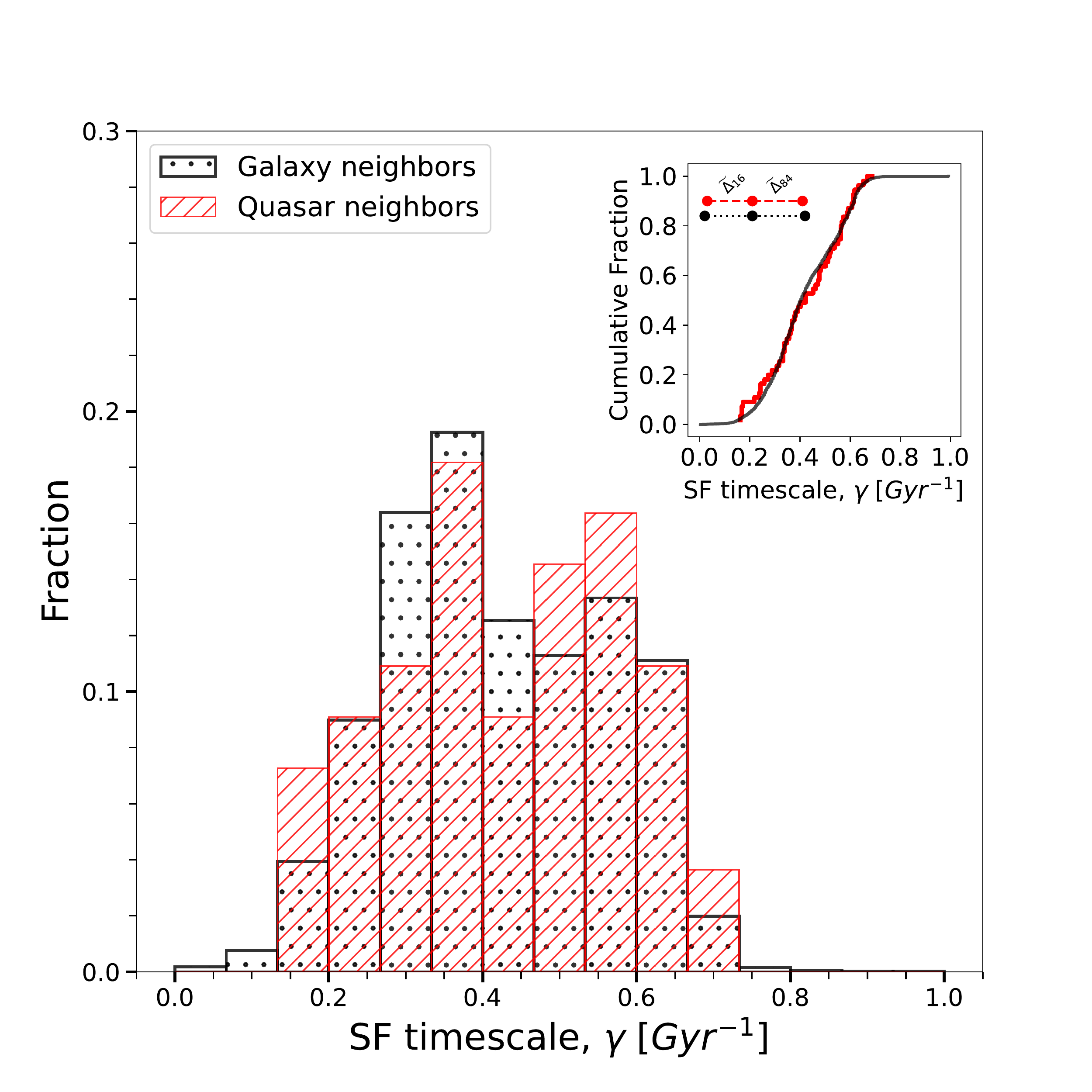}
    \includegraphics[width=0.49\textwidth]{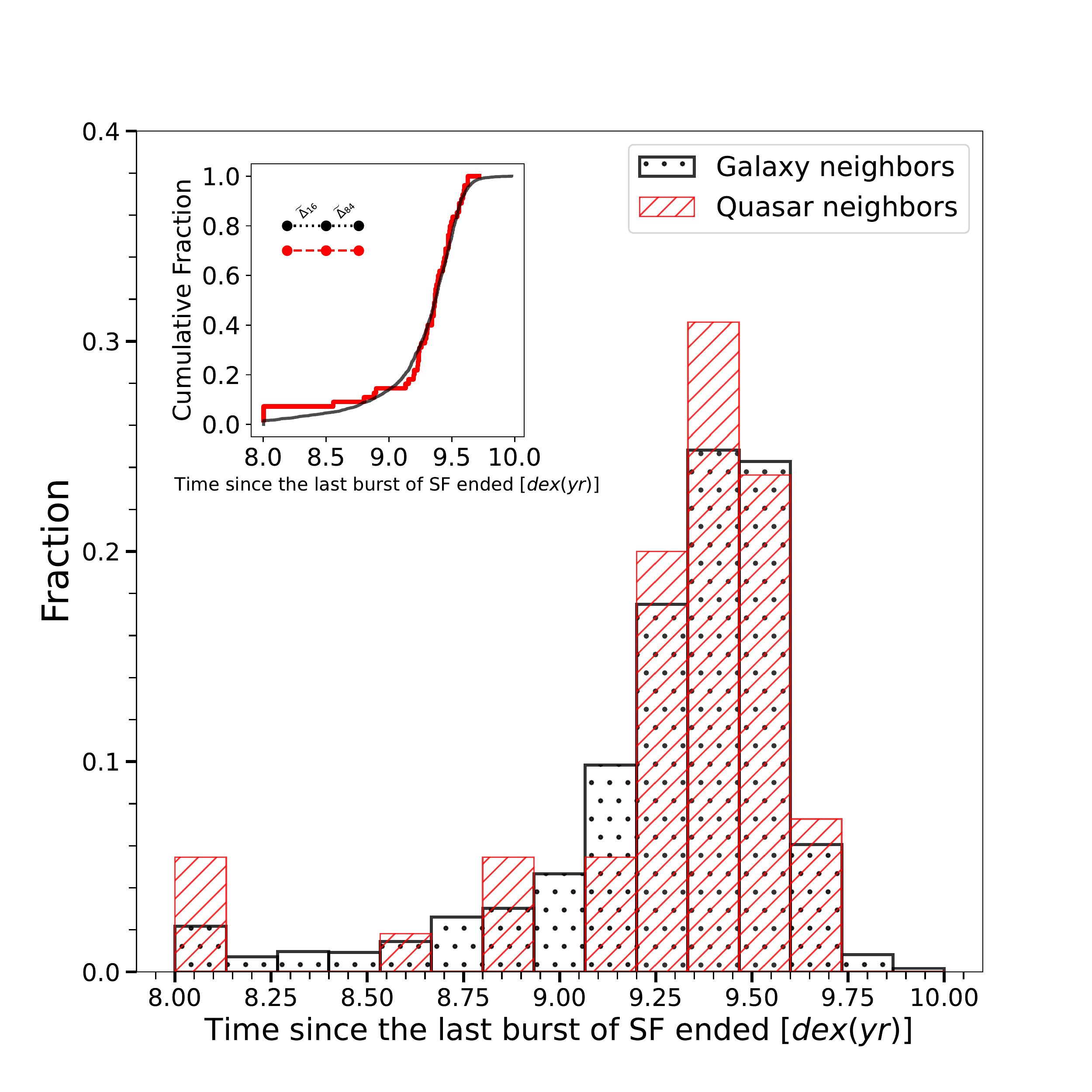}
\caption{(a) Median SF timescale. 
(b) Median time since the last burst of SF ended (quasar neighbors in red hatched bars, comparison galaxy neighbors in black dotted bars).
The inset shows the eCDF, quasar neighbors in red solid line, and comparison galaxy neighbors in black.
The typical uncertainty is represented by the median values of 16th and 84th percentiles.}
\label{fig:SF_timescale_results}
\end{figure}

\begin{figure}
\centering
    \includegraphics[width=0.47\textwidth]{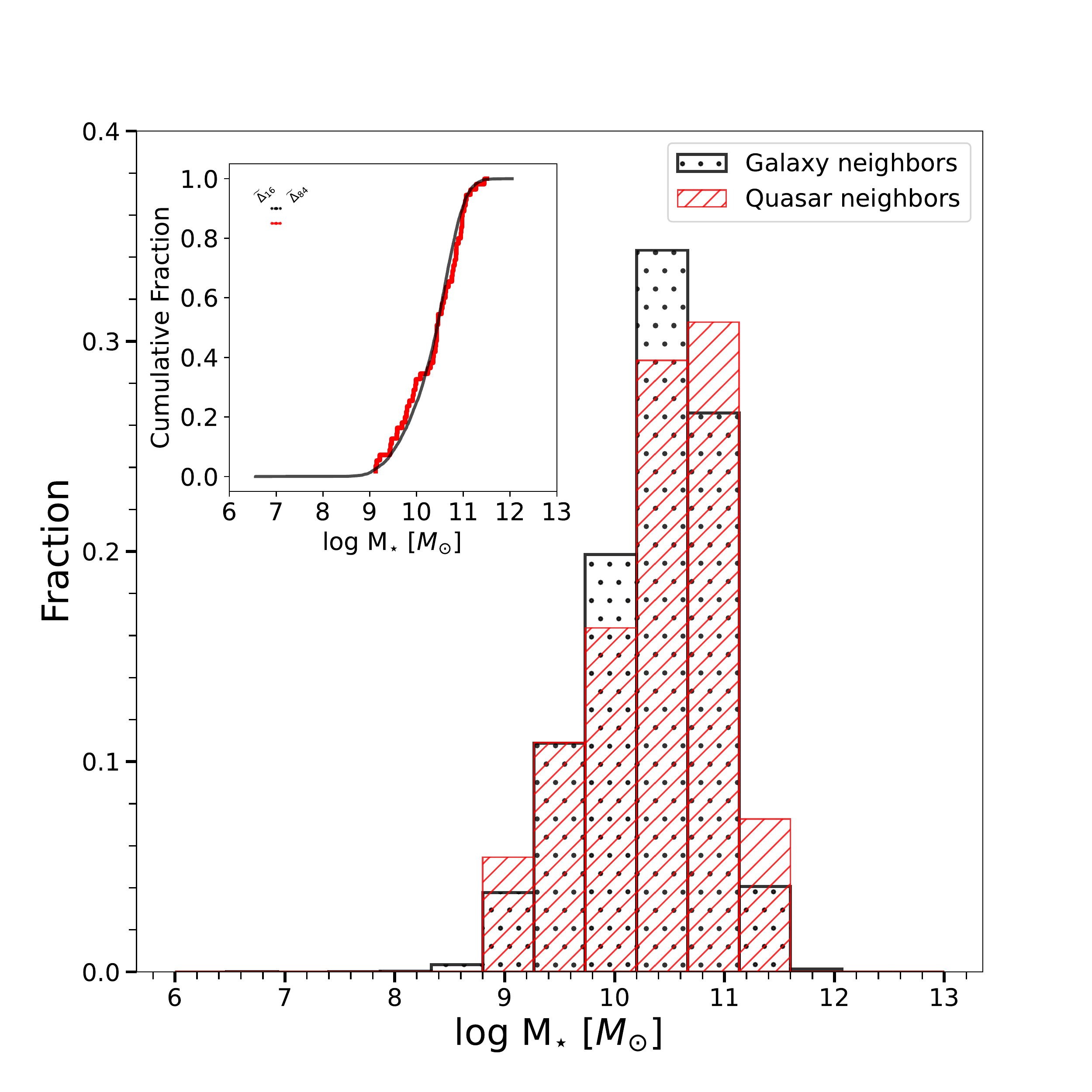}
    \includegraphics[width=0.47\textwidth]{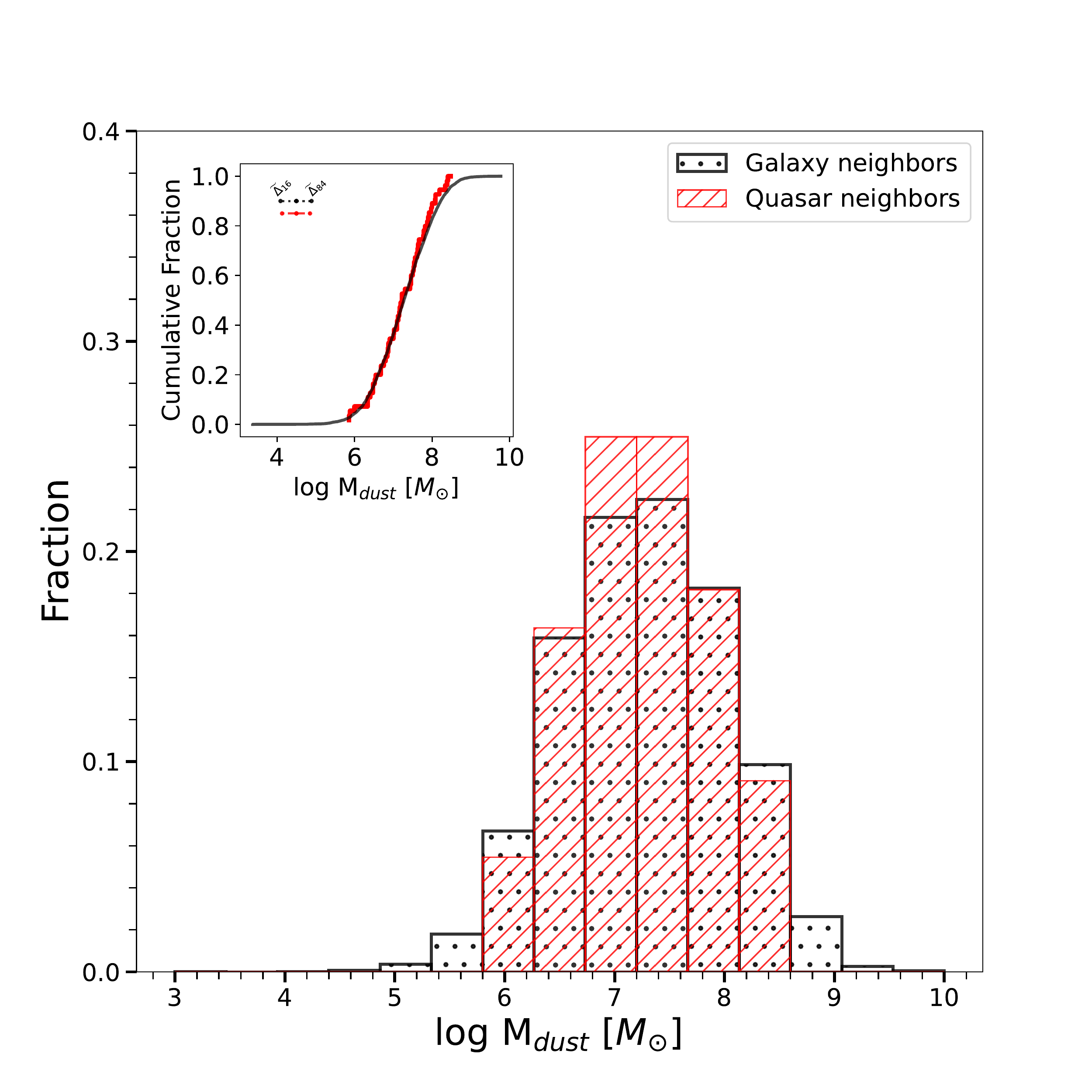}
\caption{(a) Stellar mass.
(b) Dust mass (quasar neighbors in red hatched bars, comparison galaxy neighbors in black dotted bars).
The inset shows the eCDF, quasar neighbors in red solid line, and comparison galaxy neighbors in black.
The typical uncertainty is represented by the median values of 16th and 84th percentiles.}
\label{fig:mass_results}
\end{figure}

\begin{figure}
    \centering
    \subfigure[]{\includegraphics[width=0.48\textwidth]{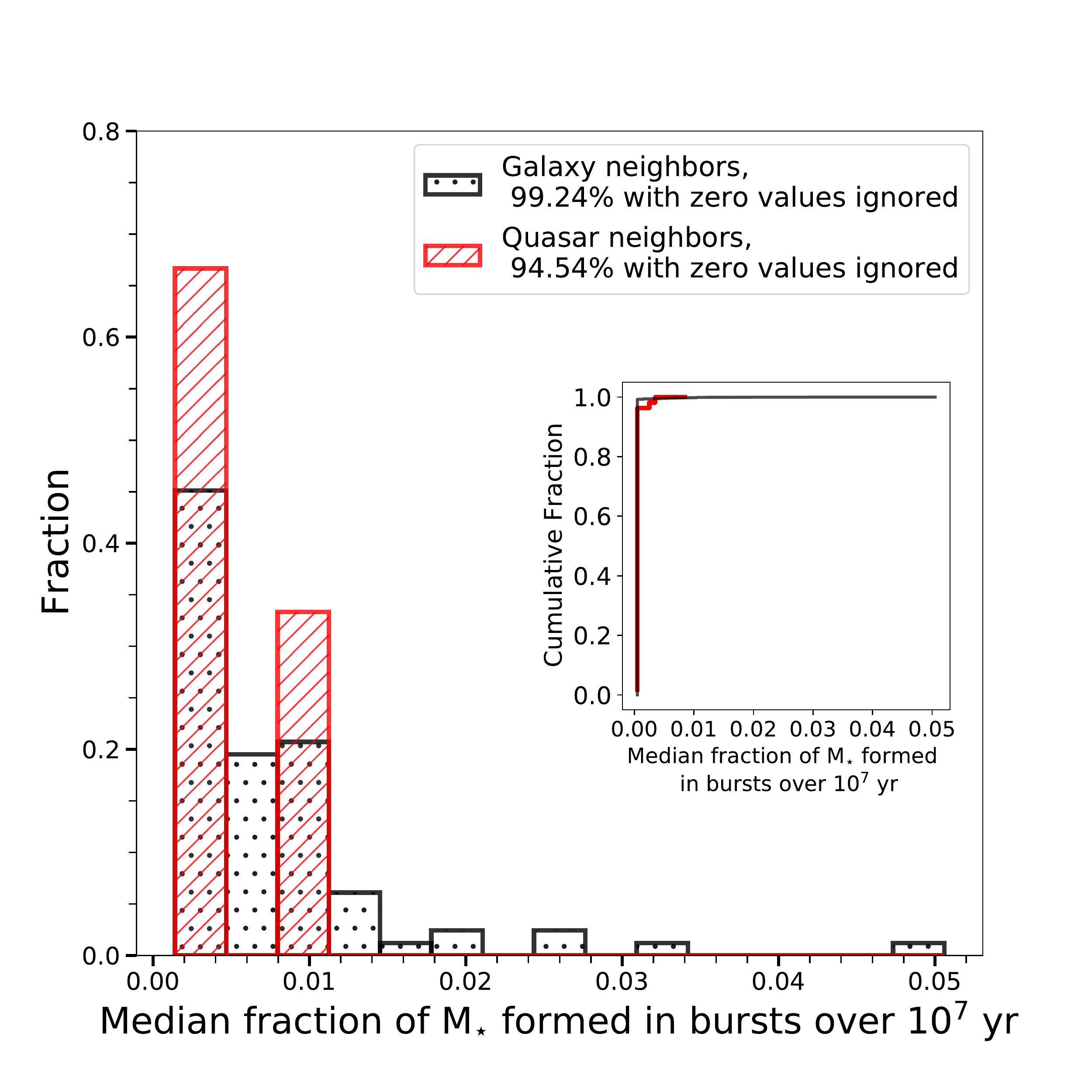}}
    \subfigure[]{\includegraphics[width=0.48\textwidth]{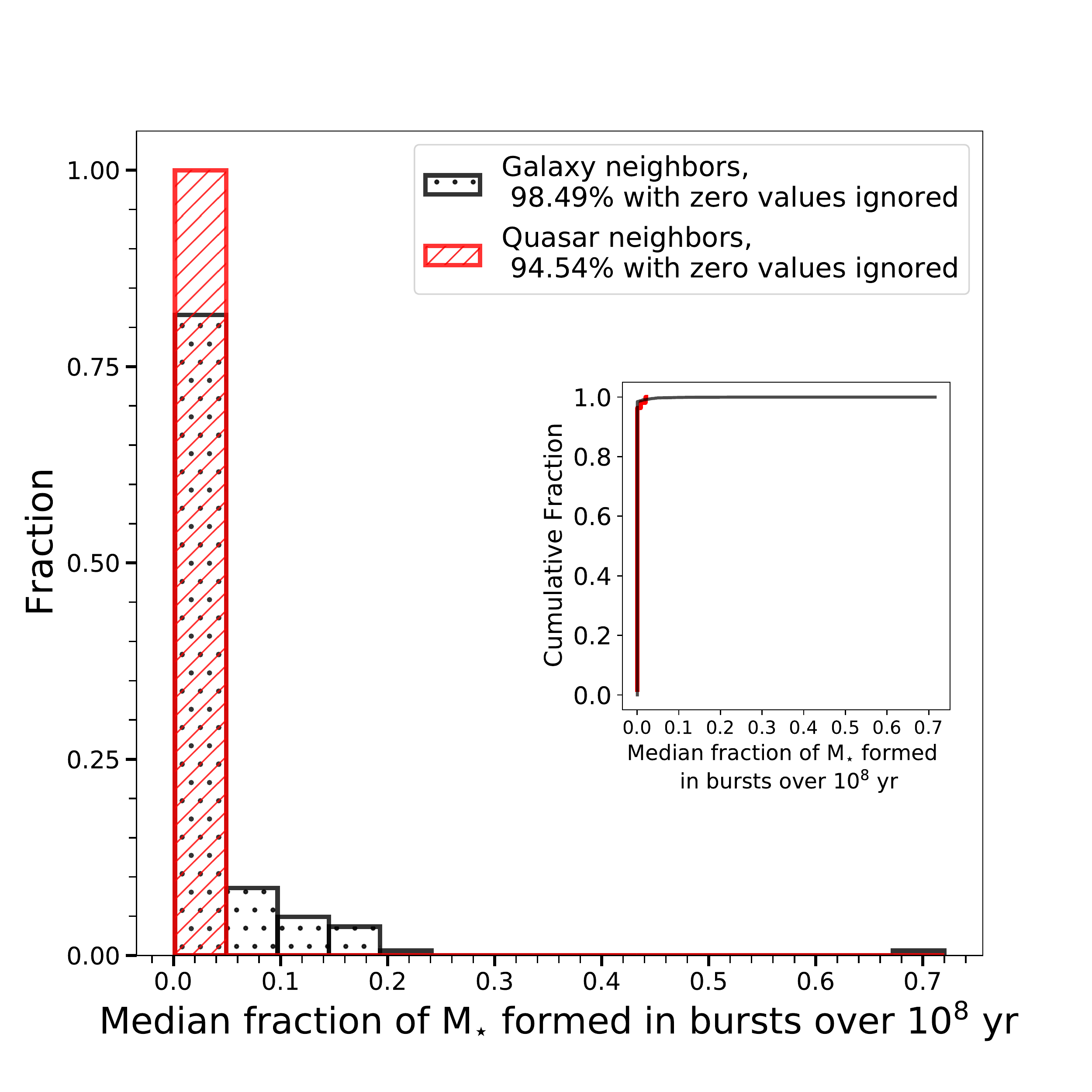}}\\
    \subfigure[]{\includegraphics[width=0.48\textwidth]{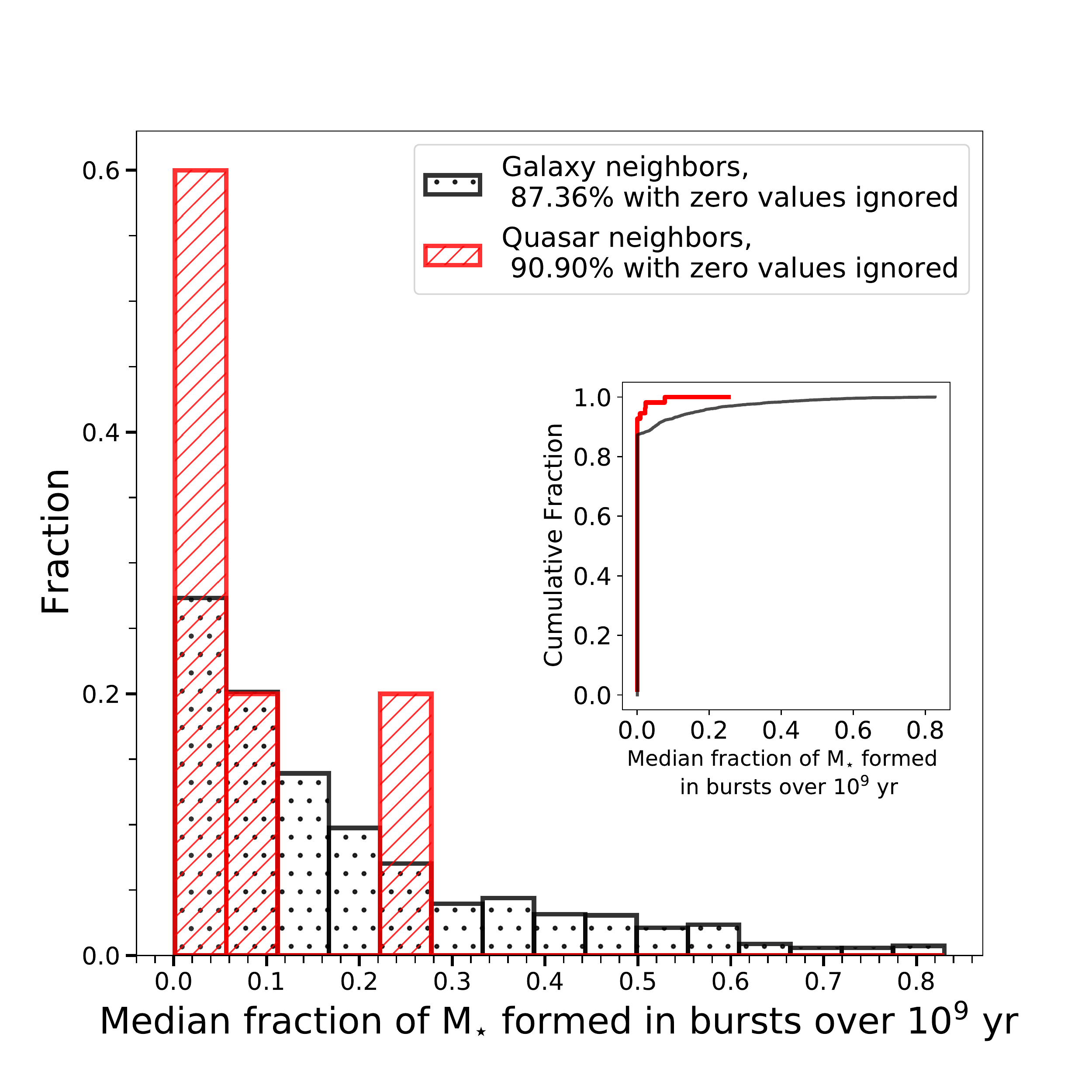}}
    \subfigure[]{\includegraphics[width=0.48\textwidth]{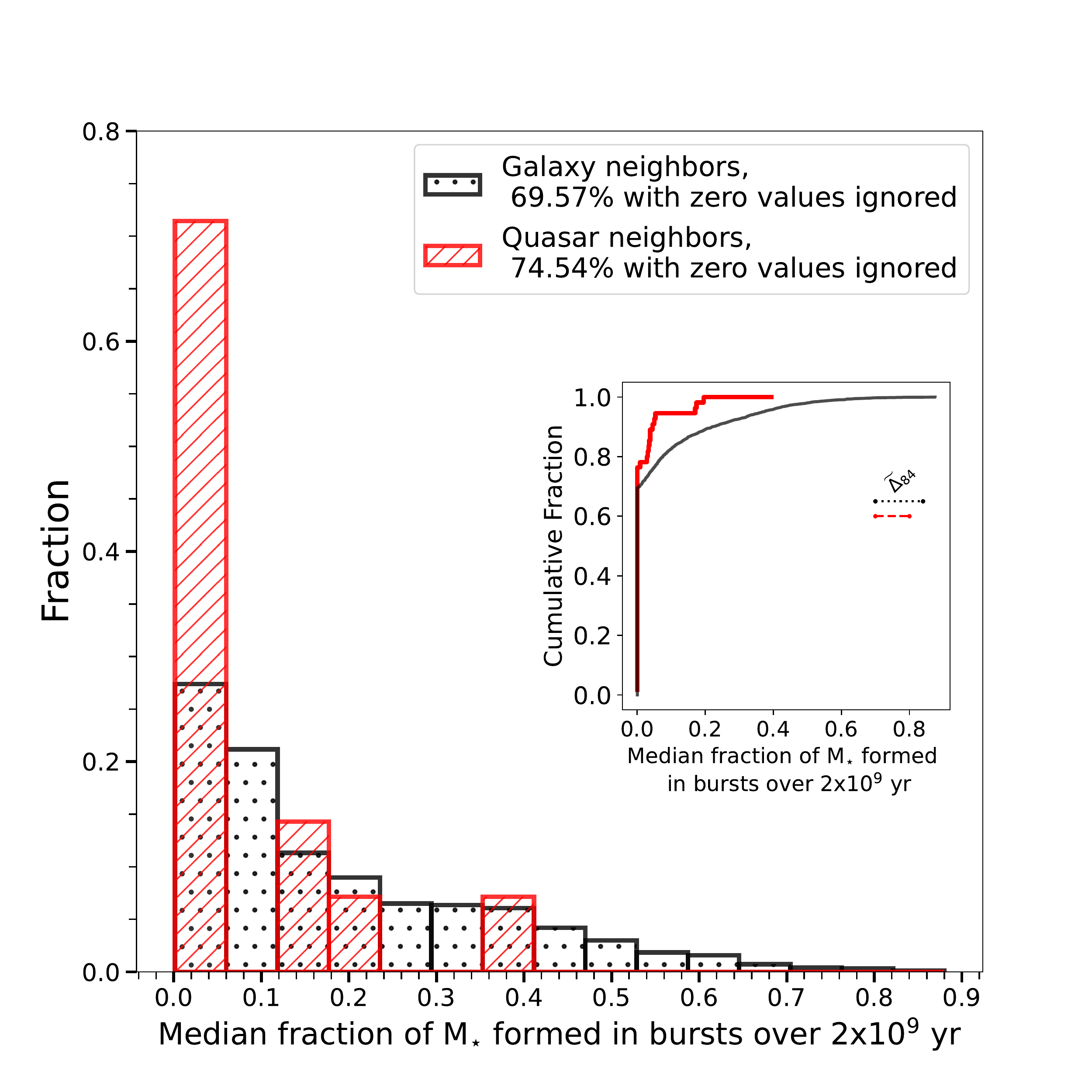}}
\caption{Median fraction of stellar mass formed in bursts over the past $10^7$ yr, $10^8$ yr, $10^9$ yr, $2 \times 10^9$ yr. 
This is the effective M$_\star$ accounting for the mass returned to the ISM (quasar neighbors in red hatched bars, comparison galaxy neighbors in black dotted bars).
The inset shows the eCDF, quasar neighbors in red solid line, and comparison galaxy neighbors in black.
The typical uncertainty is represented by the median values of 16th and 84th percentiles.
The properties in panels (a), (b), and (c) had zero-value uncertainties, while the property in panel (d) had a zero-value lower bound uncertainty.
}
\label{fig:fb17_results}
\end{figure}

\begin{figure}
\centering
    \includegraphics[width=0.45\textwidth]{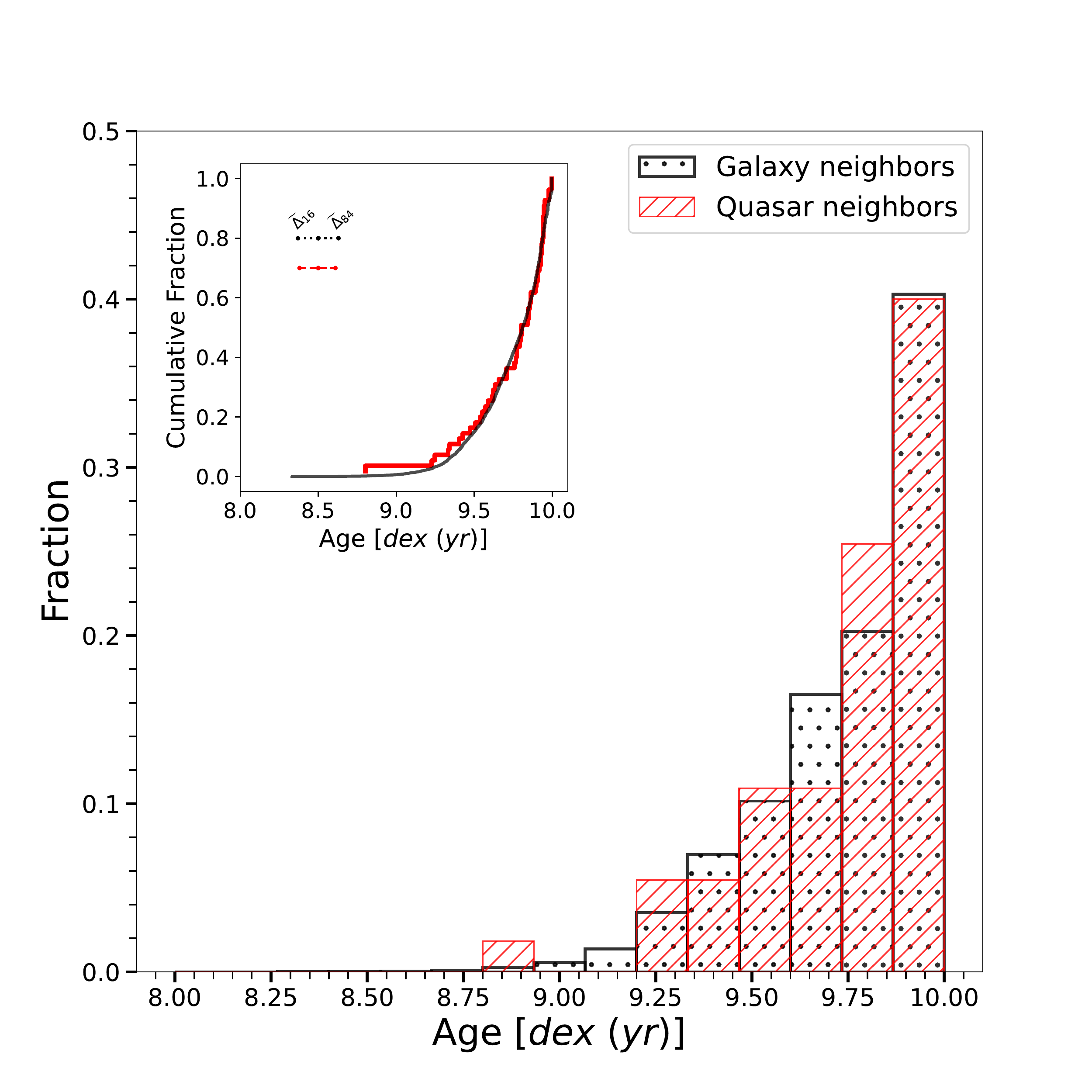}
    \includegraphics[width=0.45\textwidth]{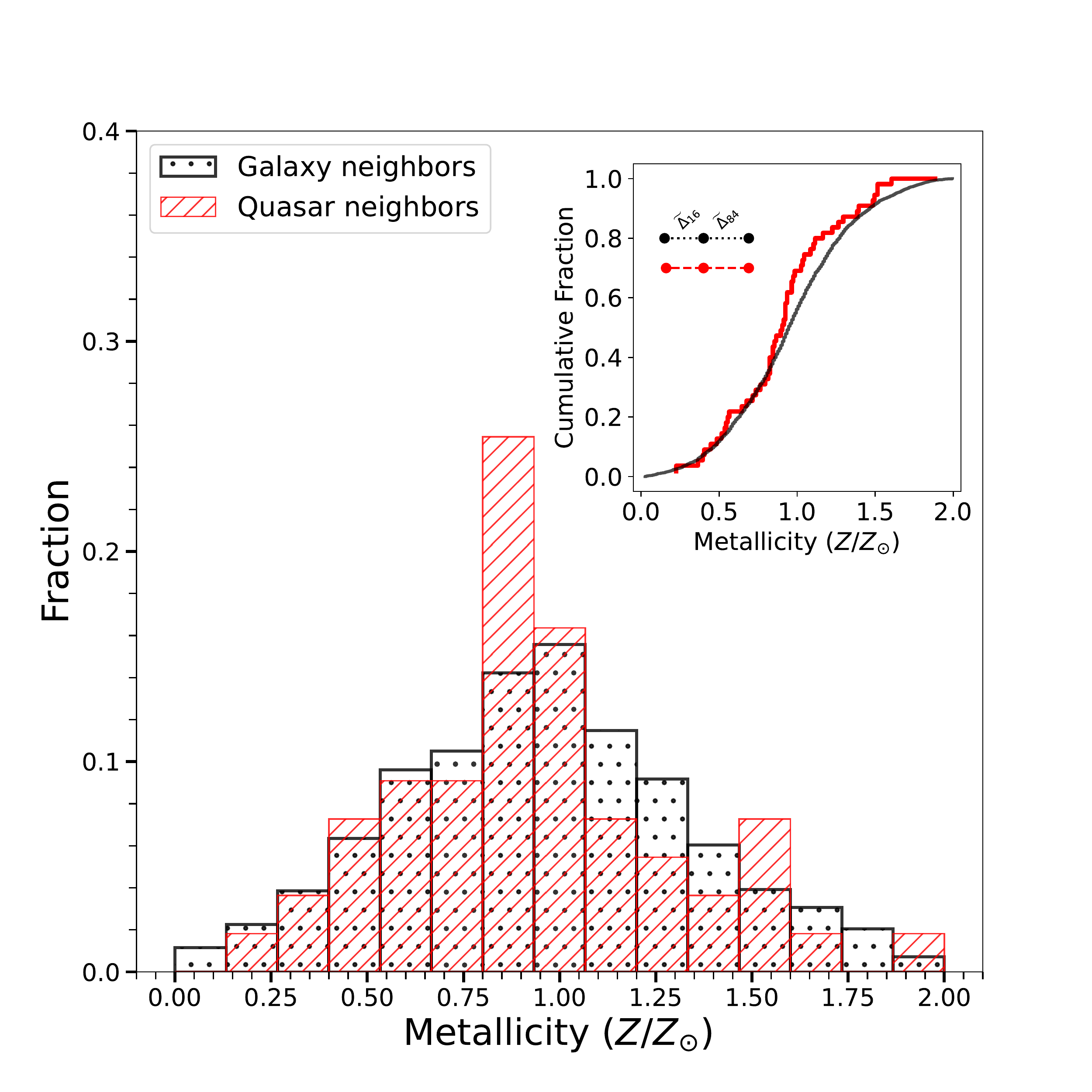}
    \caption{(a) Age.
(b) Metallicity (quasar neighbors in red hatched bars, comparison galaxy neighbors in black dotted bars).
The inset shows the eCDF, quasar neighbors in red solid line, and comparison galaxy neighbors in black.
The typical uncertainty is represented by the median values of 16th and 84th percentiles.}
\label{fig:age_results}
\end{figure}

\end{document}